\documentclass[aps,preprint]{revtex4}%
\usepackage{amsfonts}
\usepackage{amsmath}
\usepackage{amssymb}
\usepackage{graphicx}%
\begin{document}
\preprint{CTP-SCU/2015008}
\title{Black Hole Radiation with Modified Dispersion Relation in Tunneling Paradigm:
Static Frame}
\author{Jun Tao}
\email{taojun@scu.edu.cn}
\author{Peng Wang}
\email{pengw@scu.edu.cn}
\author{Haitang Yang}
\email{hyanga@scu.edu.cn}
\affiliation{Center for Theoretical Physics, College of Physical Science and Technology,
Sichuan University, Chengdu, 610064, PR China}

\begin{abstract}
To study possible deviations\ from the Hawking's prediction, we assume that
the dispersion relations of matter fields are modified at high energies and
use the Hamilton-Jacobi method to investigate the corresponding effects on the
Hawking radiation in this paper. The preferred frame is the static frame of
the black hole. The dispersion relation adopted agrees with the relativistic
one at low energies but is modified near the Planck mass $m_{p}$. We calculate
the corrections to the Hawking temperature for massive and charged particles
to $\mathcal{O}\left(  m_{p}^{-2}\right)  $ and massless and neutral particles
to all orders. Our results suggest that the thermal spectrum of radiations
near horizon is robust, e.g. corrections to the Hawking temperature are
suppressed by $m_{p}$. After the spectrum of radiations near the horizon is
obtained, we use the brick wall model to compute the thermal entropy of a
massless scalar field near the horizon of a 4D spherically symmetric black
hole. We find that the subleading logarithmic term of the entropy does not
depend on how the dispersion relations of matter fields are modified. Finally,
the luminosities of black holes are computed by using the geometric optics approximation.

\end{abstract}
\maketitle
\tableofcontents



\section{Introduction}

The classical theory of black holes predicts that anything, including light,
couldn't escape from the black holes. However, Stephen Hawking demonstrated
that quantum effects could allow black holes to radiate a thermal flux of
quantum particles \cite{IN-Hawking:1974sw}. The assumption that the ultra-high
energy modes are in their ground state was used to derive the Hawking
radiation in the framework of quantum field theory in curved spacetime. After
this discovery, it was realized that there was the trans-Planckian problem
with the calculation \cite{IN-Unruh:1976db}. Due to the exponential high
gravitational red shift near the horizon, the outgoing particles of the
Hawking radiation originate from the extremely high (e.g., trans-Planckian)
frequency modes. So the Hawking radiation relies on the validity of quantum
field theory in curved spacetime to arbitrary high energies. On the other
hand, quantum field theory is considered more like an effective field theory
of an underlying theory whose nature remains unknown \cite{IN-Weinberg:1996kw}%
. This observation poses the question of whether any unknown physics at the
Planck scale could strongly influence the Hawking radiation.

It is believed that the trans-Planckian physics manifests itself in certain
modifications of the existing models. Thus, even though a complete theory of
quantum gravity is not yet available, we can use a \textquotedblleft
bottom-to-top approach\textquotedblright\ to probe the possible effects of
quantum gravity on our current theories and experiments
\cite{IN-AmelinoCamelia:2004hm}. One possible way of how such an approach
works is via Planck-scale modifications of the usual energy-momentum
dispersion relation
\begin{equation}
p^{2}=E^{2}-m^{2}, \label{eq:uMDR}%
\end{equation}
whose possibility has been considered in the quantum-gravity literature
\cite{IN-AmelinoCamelia:1997gz,IN-Garay:1998wk,IN-AmelinoCamelia:2002wr,IN-Magueijo:2002am}%
. The modified dispersion relation (MDR) has been reviewed in the framework of
Lorentz violating theories in \cite{IN-Mattingly:2005re,IN-Liberati:2013xla}.
It has also been shown that the MDR might play a role in astronomical and
cosmological observations, such as the threshold anomalies of ultra high
energy cosmic rays and TeV photons
\cite{IN-AmelinoCamelia:1997gz,IN-Colladay:1998fq,IN-Coleman:1998ti,IN-AmelinoCamelia:2000zs,IN-Jacobson:2001tu,IN-Jacobson:2003bn}%
. Moreover, thermodynamics of black holes have been explored in the framework
of the MDR
\cite{IN-AmelinoCamelia:2004xx,IN-Ling:2005bq,IN-AmelinoCamelia:2005ik,IN-Nozari:2006ka,IN-Sefiedgar:2010we,IN-Majumder:2011xg}%
.

On the other hand, there are various methods for deriving the Hawking
radiation and calculating its temperature. Among them is a semiclassical
method of modeling Hawking radiation as a tunneling process. This method was
first proposed by Kraus and Wilczek \cite{IN-Kraus:1994by,IN-Kraus:1994fj},
which is known as the null geodesic method. They employed the dynamical
geometry approach to calculate the imaginary part of the action for the
tunneling process of s-wave emission across the horizon and related it to the
Boltzmann factor for the emission at the Hawking temperature. Later, the
tunneling behaviors of particles were investigated using the Hamilton-Jacobi
method \cite{IN-Srinivasan:1998ty,IN-Angheben:2005rm,IN-Kerner:2006vu}. In the
Hamilton-Jacobi method, one ignores the self-gravitation of emitted particles
and assumes that its action satisfies the relativistic Hamilton-Jacobi
equation. The tunneling probability for the classically forbidden trajectory
from inside to outside the horizon is obtained by using the Hamilton-Jacobi
equation to calculate the imaginary part of the action for the tunneling
process. Using the null geodesic method and Hamilton-Jacobi method, much fruit
has been achieved
\cite{IN-Hemming:2001we,IN-Medved:2002zj,IN-Vagenas:2001rm,IN-Arzano:2005rs,IN-Wu:2006pz,IN-Nadalini:2005xp,IN-Chatterjee:2007hc,IN-Akhmedova:2008dz,IN-Akhmedov:2008ru,IN-Akhmedova:2008au,IN-Banerjee:2008ry,IN-Singleton:2010gz}%
. Furthermore, the effects of quantum gravity on the Hawking radiation have
been discussed in the Hamilton-Jacobi method. In fact, the minimal length
deformed Hamilton-Jacobi equation for fermions in curved spacetime have been
introduced, and the modified Hawking temperatures have been derived
\cite{IN-Chen:2013pra,IN-Chen:2013tha,IN-Chen:2013ssa,IN-Chen:2014xsa,IN-Chen:2014xgj,IN-Mu:2015qta}%
.

In order to introduce the modified dispersion relation we need to specify one
special reference frame. The Hamilton-Jacobi equations were imported to curved
spacetime using the static preferred frame in
\cite{IN-Chen:2013pra,IN-Chen:2013tha,IN-Chen:2013ssa,IN-Chen:2014xsa,IN-Chen:2014xgj,IN-Mu:2015qta}%
, which leads us first to considering the static preferred frame in this
paper. The models with free-fall preferred frame will be investigated in
\cite{IN-Wang}. Comparisons between the results in our paper and those in
\cite{IN-Chen:2013pra,IN-Chen:2013tha,IN-Chen:2013ssa,IN-Chen:2014xsa,IN-Chen:2014xgj}
will be given at the end of the section \ref{Sec:DHJE}.

As shown in the appendix, specifying one special reference frame in the
framework of the effective field theory is just picking up a vacuum
expectation value (vev) for a vector field $u^{\mu}$. In static
frame/free-fall frame scenario, the vev of $u^{\mu}$ is the unit vector field
tangent to the static/free-fall observer's world line. In the standard model,
different vev of the Higgs field could give different results, e.g., the
results in LHC and evolution of the universe. In this sense, it is not
expected to have equivalent results from static and free-fall frame scenarios.
In fact, a brief comparison between the results of this paper and those of
free-fall scenario in \cite{IN-Wang} is given in section \ref{Sec:Con}, which
shows that differences are found for these two scenarios. Just like that the
vev of the Higgs field is determined by the results from LHC and other
experiments, the vev of $u^{\mu}$ should be given by observational and
experimental results. Before we can achieve this, every possible scenario
deserves being explored.

It is noteworthy that the modified Hawking temperature, the atmosphere entropy
in the brick wall model, and the spectrum of radiation of black holes due to
the MDR have been widely investigated. In the following, we sum up what is new
in our paper:

\begin{enumerate}
\item The modified Hawking temperature has been calculated in a heuristic
method proposed in \cite{IN-AmelinoCamelia:2004xx,IN-AmelinoCamelia:2005ik},
which was similar to that introduced by Bekenstein \cite{IN-Bekenstein:1973ur}%
. In a more rigorous way, the dispersive field theory models
\cite{IN-Unruh:1994je,IN-Brout:1995wp,IN-Corley:1996ar,IN-Corley:1997pr,IN-Himemoto:1999kd,IN-Saida:1999ap,IN-Unruh:2004zk,IN-Macher:2009tw,IN-Coutant:2011in,IN-Coutant:2014cwa,IN-Belgiorno:2014bna}
have been proposed to study the effects on the Hawking radiation due to
modifications of the dispersion relations of matter field at high energies.
These models were motivated by a hydrodynamic analogue of a black hole
radiation \cite{IN-Unruh:1994je}. Similar to the original method for deriving
the Hawking radiation, the energy fluxes for outgoing radiation were usually
obtained by calculating the Bogoliubov transformations between the initial and
final states of incoming and outgoing radiation. In these works, the Hawking
temperature was only calculated up to the leading order. In our paper, we use
the Hamilton-Jacobi method to study the dispersive field theory models and
calculate the modified effective Hawking temperature beyond the leading order.

\item The modified Stefan-Boltzmann law and luminosities of black holes due to
the MDR were discussed in \cite{IN-AmelinoCamelia:2005ik}. We also calculate
the modified luminosities of black holes in our paper. The\ geometric optics
approximation are used in \cite{IN-AmelinoCamelia:2005ik} and our paper. In
such approximation, a 4D Schwarzschild black hole can be described as a black
sphere of the radius $R$ and the temperature $T$. As a result, we consider
effects of the MDR on both $R$ (see $r_{\min}$ in eqns. $\left(
\ref{eq:R amd lamda min}\right)  $) and $T$ in our paper. On the other hand,
only effects of the MDR on $T$ were considered in
\cite{IN-AmelinoCamelia:2005ik}. Moreover, in our paper, we use detailed
balance condition to show that the average number $n_{\omega,i}$ in the mode
with the energy $\omega$ and other quantum numbers $i$ is%
\begin{equation}
n_{\omega,i}=n\left(  \frac{\omega}{T_{eff}}\right)  ,
\end{equation}
where $T_{eff}$ is the modified effective Hawking temperature depending on
$\omega$. In contrast, the authors of \cite{IN-AmelinoCamelia:2005ik} assumed
an average behavior for particles described by a unique average temperature
$T_{BH}$, which only depended on the black hole. Therefore in
\cite{IN-AmelinoCamelia:2005ik}, the average number $n_{\omega,i}$ was given
by%
\begin{equation}
n_{\omega,i}=n\left(  \frac{\omega}{T_{BH}}\right)  .
\end{equation}

\item By using the brick wall model, the atmosphere entropy of radiation of a
black hole was calculated in the framework of the MDR \cite{IN-Chang:2003sa}
and generalized uncertainty principle \cite{IN-Kim:2007if}. In these works,
only the effects of quantum gravity on the number of quantum states were
considered. In our paper, we calculate the modified atmosphere entropy in the
brick model by taking the effects of the MDR both on the number of quantum
states and Hawking temperature into account.
\end{enumerate}

The remainder of our paper is organized as follows. In section \ref{Sec:DHJE},
the deformed Hamilton-Jacobi equations incorporating the MDR are derived. In
section \ref{Sec:TOR}, we solve the deformed Hamilton-Jacobi equations to
obtain tunneling rates for massive and charged particles to $\mathcal{O}%
\left(  m_{p}^{-2}\right)  $ and massless and neutral particles to all orders.
Thermodynamics of radiations near the horizon is discussed in section
\ref{Sec:TOR}. The thermal entropy of a massless scalar field near the horizon
is computed in section \ref{Sec:EBW} by using the brick wall model. In section
\ref{Sec:BHE}, we calculate luminosities of a 4D spherically symmetric black
hole with the mass $M\gg m_{p}$ and a 2D one. Section \ref{Sec:Con} is devoted
to our discussion and conclusion, where the limitations of our calculations
are discussed. Effective field\ theories incorporating the MDR are constructed
in the appendix to obtain the deformed Hamilton-Jacobi equations. Throughout
the paper we take Geometrized units $c=G=1$, where the Planck constant $\hbar$
is square of the Planck Mass $m_{p}$.

\section{Deformed Hamilton-Jacobi Equation}

\label{Sec:DHJE}

In most cases, the MDR could take the form of%
\begin{equation}
p^{2}=m_{p}^{2}H\left(  \frac{E}{m_{p}},\frac{m}{m_{p}}\right)  ,
\label{eq:ExactMDR}%
\end{equation}
where $m_{p}$ is Planck mass and $H\left(  x,y\right)  =x^{2}-y^{2}$ for the
unmodified dispersion relation. Taylor expanding the right-hand side of eqn.
$\left(  \ref{eq:ExactMDR}\right)  $ for $E,m\ll m_{p}$ gives%
\begin{equation}
p^{2}=\sum_{i,j=0}^{\infty}h_{i,j}\frac{E^{i}m^{i}}{m_{p}^{i+j-2}},
\label{eq:MDRExpansionH}%
\end{equation}
where $h_{i,j}$ is the coefficient of $x^{i}y^{j}$ in the Taylor series of
$H\left(  x,y\right)  $ evaluated at $\left(  0,0\right)  $. Since eqn.
$\left(  \ref{eq:MDRExpansionH}\right)  $ has to become eqn. $\left(
\ref{eq:uMDR}\right)  $ when $m_{p}\rightarrow\infty$, we find
\begin{equation}
h_{0,0}=h_{0,1}=h_{1,0}=h_{1,1}=0\text{ and }h_{2,0}=h_{0,2}=1\text{.}%
\end{equation}
After some manipulations, eqn. $\left(  \ref{eq:MDRExpansionH}\right)  $ can
be put in the form of%
\begin{equation}
p^{2}=\alpha\left(  \frac{m}{m_{p}}\right)  E^{2}-\beta\left(  \frac{m}{m_{p}%
}\right)  m^{2}+\gamma\left(  \frac{m}{m_{p}}\right)  mE+\sum_{n\geq3}%
\frac{C_{n}\left(  \frac{m}{m_{p}}\right)  E^{n}}{m_{p}^{n-2}},
\label{eq:MDRExpansion}%
\end{equation}
where $\alpha\left(  0\right)  =\beta\left(  0\right)  =1$ and $\gamma\left(
0\right)  =0$. \ If the modifications to the dispersion relation are
suppressed by some the scale of Lorentz violation $\Lambda,$ the naturalness
in effective field theories would imply that $C_{n}\sim\left(  \frac{m_{p}%
}{\Lambda}\right)  ^{n}$. For\ $\Lambda\ll m_{p}$, $C_{n}$ could become much
large. To include a broader class, we consider a static black hole in the
possible presence of electromagnetic potential $A_{\mu}$ with the line element%
\begin{equation}
ds^{2}=f\left(  r\right)  dt^{2}-\frac{1}{f\left(  r\right)  }dr^{2}-C\left(
r^{2}\right)  h_{ab}\left(  x\right)  dx^{a}dx^{b}, \label{eq:BHmetric}%
\end{equation}
where $f\left(  r\right)  $ has a simple zero at $r=r_{h}$ with $f^{\prime
}\left(  r_{h}\right)  $ being finite and nonzero. The vanishing of $f\left(
r\right)  $ at point $r=r_{h}$ indicates the presence of an event horizon. We
also assume that the vector potential $A_{\mu}$ is given by
\begin{equation}
A_{\mu}=A_{t}\left(  r\right)  \delta_{\mu t}, \label{eq:vector-oneform}%
\end{equation}
which is true for charged static black holes in most cases.

The MDR breaks the Lorentz invariance in flat spacetime. Thus, one needs to
pick up a preferred frame to determine the form of the MDR. The energy and
momentum in eqn. $\left(  \ref{eq:MDRExpansion}\right)  $ are defined with
respect to the preferred frame, where can be described by the unit vector
$u^{\mu}$ tangent to the observers' world lines. Explicitly, we have
\begin{align}
E  &  =p_{\mu}u^{\mu},\nonumber\\
p^{2}  &  =E^{2}-p_{\mu}p^{\mu}, \label{eq:energy and momentum}%
\end{align}
where $p_{\mu}$ is the energy-momentum vector and $E$ and $p$ are the energy
and the norm of the momentum measured in the preferred reference frame,
respectively. When introducing the MDR into curved spacetime, we use the
vector field $u^{\mu}\left(  x_{\nu}\right)  $. To obtain the deformed
Hamilton-Jacobi equation incorporating the MDR, it is necessary to specify the
profile of the preferred frame in the black hole spacetime. One of natural
frames is a static frame hovering above the black hole. For such a frame, the
vector field $u^{\mu}\left(  x_{\nu}\right)  $ is
\begin{equation}
u^{\mu}\left(  x_{\nu}\right)  =\left(  \sqrt{g^{tt}},\vec{0}\right)  =\left(
\frac{1}{\sqrt{f\left(  r\right)  }},\vec{0}\right)  . \label{eq:4velocity}%
\end{equation}
Plugging eqn. $\left(  \ref{eq:4velocity}\right)  $ into eqns. $\left(
\ref{eq:energy and momentum}\right)  $, one finds that the energy and the
magnitude of the momentum becomes%
\begin{align}
E  &  =\frac{p_{t}}{\sqrt{f\left(  r\right)  }},\nonumber\\
p^{2}  &  =f\left(  r\right)  p_{r}^{2}+\frac{h^{ab}\left(  x\right)
}{C\left(  r^{2}\right)  }p_{a}p_{b}. \label{eq:energy-momentum-BH}%
\end{align}
It can be shown that, if the classical action $I$ is a solution of the
Hamilton-Jacobi equation, then the transformation equations give%
\begin{equation}
p_{\mu}=-\partial_{\mu}I, \label{eq:transformation-eqn}%
\end{equation}
where $-$ appears since $p_{\mu}=\left(  E,-\vec{p}\right)  $ in our metric
signature. Furthermore, since $\partial_{t}$ is a Killing vector of the
background spacetime, $\left(  \partial_{t}\right)  ^{\mu}p_{\mu}=p_{t}$ is a
constant. In fact, $p_{t}$ is the conserved energy of the particle, and we
define $\omega\equiv$ $p_{t}=-\partial_{t}I$, which means we can separate $t$
from other variables. Relating $I$ to $p_{\mu}$ via eqn. $\left(
\ref{eq:transformation-eqn}\right)  $ and putting eqns. $\left(
\ref{eq:energy-momentum-BH}\right)  $ into eqn. $\left(  \ref{eq:MDRExpansion}%
\right)  $ give the deformed Hamilton-Jacobi equation. In the appendix, the
deformed Hamilton-Jacobi equation is also derived in a more rigorous way,
specifically in the language of the effective field theory. We show there that
if a scalar/fermion obeys the MDR given in eqn. $\left(  \ref{eq:MDRExpansion}%
\right)  $ in flat spacetime, the deformed scalar/fermionic Hamilton-Jacobi
equation with respect to the preferred static frame in the black hole
background spacetime can be both written as%
\begin{equation}
X^{2}=\alpha\left(  \frac{m}{m_{p}}\right)  T^{2}-\beta\left(  \frac{m}{m_{p}%
}\right)  m^{2}+\gamma\left(  \frac{m}{m_{p}}\right)  mT+\sum_{n\geq3}%
\frac{C_{n}\left(  \frac{m}{m_{p}}\right)  T^{n}}{m_{p}^{n-2}},
\label{eq:deformedHJeqn}%
\end{equation}
where we define%
\begin{equation}
T=-\frac{\partial_{t}I+qA_{t}}{\sqrt{f\left(  r\right)  }},X^{2}=f\left(
r\right)  \left(  \partial_{r}I\right)  ^{2}+\frac{h^{ab}\left(  x\right)
\partial_{a}I\partial_{b}I}{C\left(  r^{2}\right)  }, \label{eq:TandX}%
\end{equation}
$A_{\mu\text{ }}$is the black hole's electromagnetic potential and $q$ is the
particle's charge.

\section{Tunneling Rate}

\label{Sec:TR}

In this section, we use the Hamilton-Jacobi method to investigate the
particles' tunneling across the event horizon $r=r_{h}$ of the metric $\left(
\ref{eq:BHmetric}\right)  $ by solving eqn. $\left(  \ref{eq:deformedHJeqn}%
\right)  $. Taking into account $\partial_{t}I=-\omega$, we can employ the
following ansatz for the action $I$%
\begin{equation}
I=-\omega t+W\left(  r\right)  +\Theta\left(  x\right)  ,
\end{equation}
where $\omega$ is the particle's energy. Plugging the ansatz into eqn.
$\left(  \ref{eq:TandX}\right)  $, we have%
\begin{align}
T  &  =\frac{\omega-qA_{t}\left(  r\right)  }{\sqrt{f\left(  r\right)  }%
},\nonumber\\
X^{2}  &  =f\left(  r\right)  \left[  \partial_{r}W\left(  r\right)  \right]
^{2}+\frac{h^{ab}\left(  x\right)  \partial_{a}\Theta\left(  x\right)
\partial_{b}\Theta\left(  x\right)  }{C\left(  r^{2}\right)  }.
\end{align}
The method of separation of variables gives the differential equation for
$\Theta\left(  x\right)  $%
\begin{equation}
h^{ab}\left(  x\right)  \partial_{a}\Theta\left(  x\right)  \partial_{b}%
\Theta\left(  x\right)  =\lambda, \label{eq:HJh}%
\end{equation}
where is $\lambda$ is a constant and determined by $h^{ab}\left(  x\right)  $.
Thus, one has%
\begin{equation}
X^{2}=f\left(  r\right)  \left[  \partial_{r}W\left(  r\right)  \right]
^{2}+\frac{\lambda}{C\left(  r^{2}\right)  },
\end{equation}
and eqn. $\left(  \ref{eq:deformedHJeqn}\right)  $ becomes an ordinary
differential equation for $W\left(  r\right)  $. In this section, we solve
eqn. $\left(  \ref{eq:deformedHJeqn}\right)  $ for $\partial_{r}W\left(
r\right)  $, calculate its residue at $r=r_{h}$ and find the imaginary part of
$I$ which gives the tunneling rate $\Gamma$ across the event horizon. We will
calculate $\operatorname{Im}W$ for two cases, a massive and charged particle
to $\mathcal{O}\left(  m_{p}^{-2}\right)  $ and a neutral and massless
particle to all orders.

\subsection{ Massive and Charged Particle to $\mathcal{O}\left(  m_{p}%
^{-2}\right)  $}

Consider a particle with the mass $m$ and the charge $q$. Solving eqn.
$\left(  \ref{eq:deformedHJeqn}\right)  $ for $p_{r}\equiv\partial_{r}W\left(
r\right)  $ gives
\begin{align}
p_{r}^{_{\pm}}  &  =\frac{\pm1}{\sqrt{f\left(  r\right)  }}\left(
\alpha\left(  \frac{m}{m_{p}}\right)  \frac{\tilde{\omega}^{2}\left(
r\right)  }{f\left(  r\right)  }-\beta\left(  \frac{m}{m_{p}}\right)
m^{2}-\frac{\lambda}{C\left(  r^{2}\right)  }+m\gamma\left(  \frac{m}{m_{p}%
}\right)  \frac{\tilde{\omega}\left(  r\right)  }{\sqrt{f\left(  r\right)  }%
}\right)  ^{\frac{1}{2}}\nonumber\\
&  \left(  1+\frac{1}{\alpha\left(  \frac{m}{m_{p}}\right)  \frac
{\tilde{\omega}^{2}\left(  r\right)  }{f\left(  r\right)  }-\beta\left(
\frac{m}{m_{p}}\right)  m^{2}-\frac{\lambda}{C\left(  r^{2}\right)  }%
+m\gamma\left(  \frac{m}{m_{p}}\right)  \frac{\tilde{\omega}\left(  r\right)
}{\sqrt{f\left(  r\right)  }}}%
{\displaystyle\sum\limits_{n\geq3}}
\frac{C_{n}\left(  \frac{m}{m_{p}}\right)  }{m_{p}^{n-2}}\frac{\tilde{\omega
}^{n}\left(  r\right)  }{f\left(  r\right)  ^{\frac{n}{2}}}\right)  ^{\frac
{1}{2}}, \label{eq:pr}%
\end{align}
where $+$/$-$ denotes the outgoing/ingoing solutions, and $\tilde{\omega
}\left(  r\right)  \equiv\omega-qA_{t}\left(  r\right)  $. Here, we have a
pole at $r=r_{h}$. \footnotetext[1]{This procedure will be discussed in detail
later in this section. \label{footnote}}Using the residue theory for the semi
circle$^{\left[  \ref{footnote}\right]  }$, we get%
\begin{equation}
\operatorname{Im}W_{\pm}\left(  r\right)  =\pm\frac{\sqrt{\alpha}\tilde
{\omega}\left(  r_{h}\right)  \pi}{2\kappa}\left(  1+\Delta_{qm}%
+\mathcal{O}\left(  m_{p}^{-3}\right)  \right)  , \label{eq:ImW}%
\end{equation}
where we define $\kappa=f^{\prime}\left(  r_{h}\right)  /2$ and
\begin{align}
\Delta_{qm}  &  =-\frac{C_{3}m\gamma}{4m_{p}\alpha^{2}}+\frac{1}{32m_{p}%
^{2}\kappa\alpha^{4}}\left[  \left(  24C_{4}\alpha-6C_{3}^{2}\right)
\tilde{\omega}^{\prime}\left(  r_{h}\right)  \tilde{\omega}\left(
r_{h}\right)  \alpha^{2}+\kappa m^{2}\gamma^{2}\frac{12C_{4}\alpha-15C_{3}%
^{2}}{2}\right. \nonumber\\
&  \left.  -\kappa\alpha\left(  6C_{3}^{2}-8C_{4}\alpha\right)  \left(
\frac{\lambda}{C\left(  r_{h}^{2}\right)  }+\beta m^{2}\right)  +\alpha
^{2}\left(  C_{3}^{2}-4C_{4}\alpha\right)  \frac{f^{\prime\prime}\left(
r_{h}\right)  \tilde{\omega}^{2}\left(  r_{h}\right)  }{\kappa}\right]  .
\label{eq:delta}%
\end{align}
The argument $\frac{m}{m_{p}}$ is suppressed for $\alpha\left(  \frac{m}%
{m_{p}}\right)  $, $\beta\left(  \frac{m}{m_{p}}\right)  $, $\gamma\left(
\frac{m}{m_{p}}\right)  $ and $C_{n}\left(  \frac{m}{m_{p}}\right)  $ in eqns.
$\left(  \ref{eq:ImW}\right)  $ and $\left(  \ref{eq:delta}\right)  .$

\subsection{Massless and Neutral Particle to All Orders}

We now work with a particle with $m=0$ and $q=0$. Solving eqn. $\left(
\ref{eq:deformedHJeqn}\right)  $ for $p_{r}$ gives%
\begin{equation}
p_{r}^{_{\pm}}=\frac{\pm1}{\sqrt{f\left(  r\right)  }}\left(  \frac{\omega
^{2}}{f\left(  r\right)  }-\frac{\lambda}{C\left(  r^{2}\right)  }\right)
^{\frac{1}{2}}\left(  1+\frac{1}{\frac{\omega^{2}}{f\left(  r\right)  }%
-\frac{\lambda}{C\left(  r^{2}\right)  }}%
{\displaystyle\sum\limits_{n\geq3}}
\frac{C_{n}}{m_{p}^{n-2}}\frac{\omega^{n}}{f\left(  r\right)  ^{\frac{n}{2}}%
}\right)  ^{\frac{1}{2}},
\end{equation}
where $C_{n}\equiv C_{n}\left(  0\right)  $ and we use $\alpha\left(
0\right)  =1.$ To get the residue of $p_{r}^{_{\pm}}$ at $r=r_{h}$, we first
define a few coefficients $C_{n}^{\alpha}$, $\tilde{C}_{m,n}$, and $\eta
_{l}^{k}$ as follows%
\begin{align}
\left(  1+x\right)  ^{\alpha}  &  =%
{\displaystyle\sum\limits_{n\geq0}}
C_{n}^{\alpha}x^{n},\nonumber\\
\left(
{\displaystyle\sum\limits_{n=0}^{\infty}}
C_{n+3}x\right)  ^{m}  &  =%
{\displaystyle\sum\limits_{n=0}^{\infty}}
\tilde{C}_{m,n}x^{n},\nonumber\\
\eta_{l}^{k}  &  =%
{\displaystyle\sum\limits_{m=0}^{k}}
\left(  -1\right)  ^{l}C_{m}^{\frac{1}{2}}C_{l}^{-m+\frac{1}{2}}\tilde
{C}_{m,k-m},
\end{align}
where $m$ is a non-negative integer, and $C_{n}^{\alpha}$ are generalized
binomial coefficients with $C_{n}^{\alpha}=%
{\displaystyle\prod\limits_{k=1}^{n}}
\frac{\alpha-k+1}{k}$. Therefore, one has%
\begin{align}
p_{r}^{\pm}  &  =\pm%
{\displaystyle\sum\limits_{m=0}^{\infty}}
\frac{\omega^{m}}{m_{p}^{m}f^{\frac{m}{2}}\left(  r\right)  }\frac
{C_{m}^{\frac{1}{2}}\omega}{f\left(  r\right)  }\left(  1-\frac{f\left(
r\right)  \lambda}{C\left(  r^{2}\right)  \omega^{2}}\right)  ^{-m+\frac{1}%
{2}}\left(
{\displaystyle\sum\limits_{n=0}^{\infty}}
\frac{C_{n+3}}{m_{p}^{n}}\frac{\omega^{n}}{f^{\frac{n}{2}}\left(  r\right)
}\right)  ^{m}\nonumber\\
&  =\pm\frac{\omega}{f\left(  r\right)  }%
{\displaystyle\sum\limits_{l=0}^{\infty}}
{\displaystyle\sum\limits_{k=0}^{\infty}}
\frac{\lambda^{l}}{C^{l}\left(  r^{2}\right)  }\frac{\eta_{l}^{k}}{m_{p}^{k}%
}\frac{\omega^{k-2l}}{f^{\frac{k}{2}-l}\left(  r\right)  }\nonumber\\
&  \sim\pm\frac{\omega}{f\left(  r\right)  }%
{\displaystyle\sum\limits_{l=0}^{\infty}}
{\displaystyle\sum\limits_{k=0}^{\infty}}
\frac{\lambda^{l}}{C^{l}\left(  r^{2}\right)  }\frac{\eta_{l}^{2k+2l}}%
{m_{p}^{2k+2l}}\frac{\omega^{2k}}{f^{k}\left(  r\right)  }\nonumber\\
&  \sim\pm\omega%
{\displaystyle\sum\limits_{l=0}^{\infty}}
\left[  \frac{\lambda}{C\left(  r_{h}^{2}\right)  m_{p}^{2}}\right]  ^{l}%
{\displaystyle\sum\limits_{k=0}^{\infty}}
\eta_{l}^{2k+2l}\frac{\omega^{2k}}{m_{p}^{2k}}\left(  \frac{C^{l}\left(
r_{h}^{2}\right)  }{C^{l}\left(  r^{2}\right)  }\frac{1}{f^{k+1}\left(
r\right)  }\right)  ,
\end{align}
where we only keep terms contributing to the residue and set $k\rightarrow
k+2l$ in the third line. Furthermore, we denote the residue of $\frac
{C^{l}\left(  r_{h}^{2}\right)  }{C^{l}\left(  r^{2}\right)  }\frac{1}%
{f^{k+1}\left(  r\right)  }$ at $r=r_{h}$ by
\begin{equation}
\text{Res}\left(  \frac{C^{l}\left(  r_{h}^{2}\right)  }{C^{l}\left(
r^{2}\right)  }\frac{1}{f^{k+1}\left(  r\right)  },r_{h}\right)  =\frac
{\zeta_{k}^{l}}{2\kappa}.
\end{equation}
Using the residue theory for the semi circle, one has%
\begin{equation}
\operatorname{Im}W_{\pm}\left(  r\right)  =\pm\frac{\omega\pi}{2\kappa}\left(
1+\Delta\right)  , \label{eq:ImW-massless}%
\end{equation}
where%
\begin{equation}
\Delta=%
{\displaystyle\sum\limits_{l+k\geq1}^{\infty}}
\left[  \frac{\lambda}{C\left(  r_{h}^{2}\right)  m_{p}^{2}}\right]  ^{l}%
\eta_{l}^{2k+2l}\zeta_{k}^{l}\frac{\omega^{2k}}{m_{p}^{2k}}.
\label{eq:DeltaMassless}%
\end{equation}
Note that $\eta_{0}^{0}=\zeta_{0}^{0}=1.$

\subsection{Calculating $\lambda$}

It is easy to see that $\lambda$ depends on $h_{ab}\left(  x\right)  .$ Here
we consider two kinds of black holes, 4D cylindrically and spherically
symmetric black holes. For a 4D cylindrically symmetric black hole, we have%
\begin{equation}
h_{ab}\left(  x\right)  dx^{a}dx^{b}=d\theta^{2}+\alpha^{2}dz^{2},
\end{equation}
where $-\infty<z<\infty,$ $0\leq\theta\leq2\pi$, and $\alpha$ is some
constant. Since $\partial_{\theta}$ and $\partial_{z}$ are the Killing fields
of the background spacetime, we can separate the variables and consider a
solution for eqn. $\left(  \ref{eq:HJh}\right)  $ of the form%
\begin{equation}
\Theta=J_{\theta}\theta+J_{z}z,
\end{equation}
where $J_{\theta}$ and $J_{z}$ are constant, and $J_{\theta}$ is the angular
momentum along $z$-axis. The periodicity of $\theta$ gives $J_{\theta}=n\hbar$
with $n\in Z$. Thus, one finds%
\begin{equation}
\lambda=J_{\theta}^{2}+\frac{J_{z}^{2}}{\alpha^{2}},
\end{equation}
where $J_{\theta}=n\hbar$ with $n\in Z$.

For a 4D spherically symmetric black hole, we have%
\begin{equation}
h_{ab}\left(  x\right)  dx^{a}dx^{b}=d\theta^{2}+\sin^{2}\theta d\phi^{2},
\end{equation}
where $0\leq\theta\leq\pi$ and $0\leq\phi\leq2\pi$. $\partial_{\phi}$ is the
Killing vector so we consider a solution for eqn. $\left(  \ref{eq:HJh}%
\right)  $ of the form%
\begin{equation}
\Theta=Y\left(  \theta\right)  +J_{\phi}\phi, \label{eq:theta}%
\end{equation}
where $J_{\phi}$ is the angular momentum along $z$-axis. The periodicity of
$\phi$ gives $J_{\phi}=m\hbar$ with $m\in Z$ . Since the magnitude of the
angular momentum of the particle $L$ can be expressed in terms of $p_{\theta
}\equiv\partial_{\theta}Y\left(  \theta\right)  $ and $J_{\phi},$
\begin{equation}
L^{2}=p_{\theta}^{2}+\frac{p_{\phi}^{2}}{\sin^{2}\theta},
\end{equation}
eqn. $\left(  \ref{eq:HJh}\right)  $ gives $\lambda=L^{2}$. Putting eqn.
$\left(  \ref{eq:theta}\right)  $ into eqn. $\left(  \ref{eq:HJh}\right)  $,
one gets%
\begin{equation}
p_{\theta}=\sqrt{\lambda-\frac{m^{2}\hbar^{2}}{\sin^{2}\theta}}.
\end{equation}
On the other hand, the Sommerfeld quantization for $p_{\theta}$ gives%
\begin{equation}%
{\displaystyle\oint}
d\theta p_{\theta}=2\pi\left(  n+\frac{1}{2}\right)  \hbar,
\label{eq:sommerfeldquantization}%
\end{equation}
where $n$\ is a non-negative integer. The integral in eqn. $\left(
\ref{eq:sommerfeldquantization}\right)  $ is calculated in the classically
allowed region where $p_{\theta}$ is real, which requires that $\lambda\geq
m^{2}\hbar^{2}$. Since the integral is taken over the whole period of the
classical motion, the left-hand side of eqn. $\left(
\ref{eq:sommerfeldquantization}\right)  $ is given by%
\begin{equation}%
{\displaystyle\oint}
d\theta p_{\theta}=\int_{\arcsin\frac{\left\vert m\right\vert \hbar}%
{\sqrt{\lambda}}}^{\pi-\arcsin\frac{\left\vert m\right\vert \hbar}%
{\sqrt{\lambda}}}p_{\theta}^{+}d\theta+\int_{\pi-\arcsin\frac{\left\vert
m\right\vert \hbar}{\sqrt{\lambda}}}^{\arcsin\frac{\left\vert m\right\vert
\hbar}{\sqrt{\lambda}}}p_{\theta}^{-}d\theta=2\int_{\arcsin\frac{\left\vert
m\right\vert \hbar}{\sqrt{\lambda}}}^{\pi-\arcsin\frac{\left\vert m\right\vert
\hbar}{\sqrt{\lambda}}}p_{\theta}d\theta,
\end{equation}
where $p_{\theta}^{+}=p_{\theta}$ and $p_{\theta}^{-}=-p_{\theta}$.
Integrating the quantization integral, one finds that eqn. $\left(
\ref{eq:sommerfeldquantization}\right)  $ becomes
\begin{equation}
2\pi\left(  \sqrt{\lambda}-\left\vert m\right\vert \hbar\right)  =2\pi\left(
n+\frac{1}{2}\right)  \hbar. \label{eq:lamda}%
\end{equation}
Solving eqn. $\left(  \ref{eq:lamda}\right)  $ for $\lambda$ gives the WKB
leading quantization of the angular momentum%
\begin{equation}
\lambda=\left(  l+\frac{1}{2}\right)  ^{2}\hbar^{2}, \label{eq:lamda1}%
\end{equation}
where $l=n+\left\vert m\right\vert =0,1,\cdots$ with $\left\vert m\right\vert
\leq l$. Note that the difference between the exact quantization of the
angular momentum $L^{2}=l\left(  l+1\right)  \hbar^{2}$ and the WKB leading
quantization $L^{2}=\left(  l+\frac{1}{2}\right)  ^{2}\hbar^{2}$ is
$\frac{\hbar^{2}}{4}$.

\subsection{Tunneling Rate}

When one calculates the quantum tunneling rate from $\operatorname{Im}W_{\pm}%
$, there is so called \textquotedblleft factor-two problem\textquotedblright%
\ \cite{TR-Akhmedov:2006pg}. Thus, one may have a black hole temperature which
is twice the expected result. One of solutions is proposed by Mitra
\cite{TR-Mitra:2006qa}. Mitra noted that in general, the action $I$ could
include some complex constant of integration $\mathcal{K}$. In this way, the
imaginary part of $I$ becomes%
\begin{equation}
\operatorname{Im}I_{\pm}=\operatorname{Im}W_{\pm}+\operatorname{Im}\mathcal{K}%
\end{equation}
$+$/$-$ denotes the outgoing/ingoing solutions. In the semi-classical method,
the absorption probability and the emission probability for a black hole are
given by%
\begin{align}
P_{emit}  &  \propto\exp\left(  -\frac{2}{\hbar}\operatorname{Im}I_{+}\right)
=\exp\left[  -\frac{2}{\hbar}\left(  \operatorname{Im}W_{+}+\operatorname{Im}%
\mathcal{K}\right)  \right]  ,\nonumber\\
P_{abs}  &  \propto\exp\left(  -\frac{2}{\hbar}\operatorname{Im}I_{-}\right)
=\exp\left[  -\frac{2}{\hbar}\left(  \operatorname{Im}W_{-}+\operatorname{Im}%
\mathcal{K}\right)  \right]  . \label{eq:abs}%
\end{align}
On the other hand, it is noted that the classical theory of black holes tells
us that an incoming particle is absorbed with the probability equalling to
one. Thus, one can choose $\mathcal{K}$ to impose the classical constraint on
the absorption probability, which is $\operatorname{Im}\mathcal{K=-}%
\operatorname{Im}W_{-}$. So eqn. $\left(  \ref{eq:abs}\right)  $\ gives\ that
the probability of a particle tunneling from inside to outside the horizon is%
\begin{equation}
P_{emit}\propto\exp\left[  -\frac{2}{\hbar}\left(  \operatorname{Im}%
W_{+}-\operatorname{Im}W_{-}\right)  \right]  . \label{eq:prob-emit}%
\end{equation}

Another way to circumvent this problem is considering both the contributions
from spatial and temporal parts of the action to the tunneling rates.

\textbf{Spatial Contribution:} Along an open path in phase spaces of two
canonically equivalent frames, one in general has that%
\begin{equation}
\int pdx\neq\int PdX.
\end{equation}
Therefore, $\operatorname{Im}W_{\pm}=\operatorname{Im}\int p_{r}^{\pm}dr$ used
in eqns. $\left(  \ref{eq:abs}\right)  $ are not invariant under canonical
transformations. On the other hand, a closed contour integral $%
{\displaystyle\oint}
pdx$ is invariant under canonical transformations. It was then suggested
\ \cite{TR-Akhmedov:2006pg,TR-Akhmedova:2008au,TR-Chowdhury:2006sk} that the
tunneling rates could be given by%
\begin{equation}
P_{emit/abs}\propto\exp\left(  \pm\frac{1}{\hbar}\operatorname{Im}%
{\displaystyle\oint}
p_{r}dr\right)  \text{,} \label{eq:RP}%
\end{equation}
where $%
{\displaystyle\oint}
p_{r}dr=\int p_{r}^{+}dr-\int p_{r}^{-}dr$. It showed that eqns. $\left(
\ref{eq:prob-emit}\right)  $ and $\left(  \ref{eq:RP}\right)  $ yielded
different results for the Schwarzschild spacetime in Painleve-Gulstrand
coordinates \cite{TR-Akhmedova:2008au} and thin shells from black holes
\cite{TR-Chowdhury:2006sk}. Eqn. $\left(  \ref{eq:RP}\right)  $ is preferred
over eqn. $\left(  \ref{eq:prob-emit}\right)  $ since eqn. $\left(
\ref{eq:RP}\right)  $ is a proper observable. When $p_{r}^{\pm}$ have the same
magnitude but opposite signs, eqns. $\left(  \ref{eq:prob-emit}\right)  $ and
$\left(  \ref{eq:RP}\right)  $ give the same results.

\textbf{Temporal Contribution:} As shown in
\cite{TR-Chowdhury:2006sk,TR-Akhmedova:2008dz,TR-deGill:2010nb}, the temporal
part contribution came from the "rotation" which connects the interior region
and the exterior region of the black hole. It was found in
\cite{TR-deGill:2010nb} that the direction in which the horizon was crossed
did not affect the sign of the temporal contribution. However, the sign of the
spatial contribution changed when the direction was reversed. Thus, the
temporal contributions to $P_{emit/abs}$ were the same. When the horizon was
crossed once, the action $I$ got a contribution of $\operatorname{Im}\left(
\omega\Delta t\right)  =\frac{\pi\omega}{2\kappa}$, and for a round trip, the
total contribution was $\operatorname{Im}\left(  \omega\Delta t\right)
=\frac{\pi\omega}{\kappa}.$

Taking into account the spatial and temporal contributions, one has for the
absorption probability
\begin{equation}
P_{abs}\propto\exp\left[  -\frac{1}{\hbar}\left(  -\operatorname{Im}%
{\displaystyle\oint}
p_{r}dr+\operatorname{Im}\left(  \omega\Delta t\right)  +\operatorname{Im}%
\mathcal{K}\right)  \right]  ,
\end{equation}
and for the emission probability%
\begin{equation}
P_{emit}\propto\exp\left[  -\frac{1}{\hbar}\left(  \operatorname{Im}%
{\displaystyle\oint}
p_{r}dr+\operatorname{Im}\left(  \omega\Delta t\right)  +\operatorname{Im}%
\mathcal{K}\right)  \right]  ,
\end{equation}
where $\mathcal{K}$ is a constant of integration. Imposing the classical
constraint on the absorption probability, one gets%
\begin{equation}
\operatorname{Im}\mathcal{K}=\operatorname{Im}%
{\displaystyle\oint}
p_{r}dr-\operatorname{Im}\left(  \omega\Delta t\right)  ,
\end{equation}
and
\begin{equation}
P_{emit}\propto\left[  -\frac{2}{\hbar}\left(  \operatorname{Im}%
{\displaystyle\oint}
p_{r}dr\right)  \right]  =\exp\left[  -\frac{2}{\hbar}\left(
\operatorname{Im}W_{+}-\operatorname{Im}W_{-}\right)  \right]  .
\end{equation}

Both approaches give the same expression for $P_{emit}$. There is a Boltzmann
factor in $P_{emit}$ with an effective temperature. Using eqns. $\left(
\ref{eq:ImW}\right)  $ and $\left(  \ref{eq:ImW-massless}\right)  ,$ we find
that the effective temperature for a massive and charged particle is
\begin{equation}
T_{eff}=\frac{T_{0}}{\sqrt{\alpha}\left(  1+\Delta_{qm}\right)  }%
+\mathcal{O}\left(  m_{p}^{-3}\right)  , \label{eq:TeffMassive}%
\end{equation}
and, that for a massless and neutral particle is%
\begin{equation}
T_{eff}=\frac{T_{0}}{1+\Delta}, \label{eq:TeffMassless}%
\end{equation}
where we define $T_{0}=\frac{\hbar\kappa}{2\pi}$ and take $k_{B}=1$.

\subsection{Discussion}

In this section, we suppose that outgoing particles tunnel from $r_{1}<r_{h}$
to $r_{2}>r_{h}$ while ingoing particles from $r_{2}$ to $r_{1}$. To to obtain
the imaginary part of $I$ for the tunneling process, we have to give an
prescription for evaluating the integrals of $\operatorname{Im}W_{\pm}%
=\int_{r_{1}}^{r_{2}}p_{r}^{\pm}dr$. Following the Feynman's $i\epsilon
$--prescription \cite{TR-Vanzo:2011wq}, we take the contour of the integral to
be an infinitesimal semicircle below the pole at $r=r_{h}$ for outgoing
particles. Thus, the integral becomes%
\begin{equation}
\int_{r_{1}}^{r_{2}}p_{r}^{+}dr=\int_{r_{1}}^{r_{h}-\varepsilon}p_{r}%
^{+}dr+\int_{C_{B,\varepsilon}}p_{r}^{+}dr+\int_{r_{h}+\varepsilon}^{r_{2}%
}p_{r}^{+}dr,
\end{equation}
where we denote the semicircle centered at $r=r_{h}$ with the radius of $R$
going from $r_{h}+R/r_{h}-R$ to $r_{h}-R/r_{h}+R$ in the upper/lower half
complex plane by $C_{U/B,R}$. Since the contributions from the ranges $\left(
r_{1},r_{h}-\varepsilon\right)  $ and $\left(  r_{h}+\varepsilon,r_{2}\right)
$ are real, the imaginary part of $W_{+}$ is%
\begin{equation}
\operatorname{Im}W_{+}=\operatorname{Im}\int_{C_{B,\varepsilon}}p_{r}^{+}dr.
\end{equation}
Similarly, one has for ingoing particles%
\begin{equation}
\operatorname{Im}W_{-}=\operatorname{Im}\int_{C_{U,\varepsilon}}p_{r}^{-}dr.
\end{equation}

To get $\operatorname{Im}W_{\pm}$, we expand $p_{r}^{\pm}$ in powers of
$\frac{\omega}{\sqrt{f\left(  r\right)  }m_{p}}$%
\begin{equation}
p_{r}^{\pm}=\sum_{n\geq0}\frac{\omega^{n+1}p_{n}^{\pm}\left(  r\right)
}{f^{\frac{n}{2}+1}\left(  r\right)  m_{p}^{n}}, \label{eq:pr-expansion}%
\end{equation}
where $p_{n}^{\pm}\left(  r\right)  $ are some analytic functions of $r$
around $r=r_{h}$. For the usual case, only the first term in eqn. $\left(
\ref{eq:pr-expansion}\right)  $ appears. Thus, we can Laurent expand $f\left(
r\right)  $\ with respect to $r$ at $r=r_{h}$ to evaluate $\int
_{C_{B/U,\varepsilon}}p_{r}^{\pm}dr$ as $\varepsilon\rightarrow0$. However,
these expansions for $p_{r}^{\pm}$ in the cases incorporating the MDR look
suspicious on $C_{U/B,\varepsilon}$ as $\varepsilon\rightarrow0$. In fact,
$\frac{\omega}{\sqrt{f\left(  r\right)  }m_{p}}$ can become larger than $1$ if
$r$ is close enough to $r_{h}$. Thus, we can not trust the expansions for
$p_{r}^{\pm}$ any more on $C_{U/B,\varepsilon}$. Nevertheless, we can assume
that the singularity structure of $p_{r}$ in the MDR cases is the same as that
in the usual case except the order of the pole at $r=r_{h}$. This assumption
means the MDR effects do not introduce branch cuts or new poles for $p_{r}$ in
the upper or lower half complex plane. Note that one may need a complete
theory of quantum gravity to justify the assumption. Now consider the
semicircles $C_{U/B,R}$ with large enough $R$, which lies in the region where
the expansion for $p_{r}$ can be trusted. Under the assumption, there are no
poles inside the area enclosed by $\left(  R,r_{h}-\varepsilon\right)  $,
$C_{U/B,\varepsilon}$, $\left(  r_{h}+\varepsilon,R\right)  $, and $C_{U/B,R}%
$. Thus, we have%
\begin{align}
\operatorname{Im}W_{\pm}  &  =\operatorname{Im}\int_{r_{1}}^{r_{h}-R}%
p_{r}^{\pm}dr+\operatorname{Im}\int_{C_{B/U,R}}p_{r}^{\pm}dr+\operatorname{Im}%
\int_{r_{h}+R}^{r_{2}}p_{r}^{\pm}dr\nonumber\\
&  =\operatorname{Im}\int_{C_{B/U,R}}p_{r}^{\pm}dr=\sum_{n\geq0}\frac
{\omega^{n+1}}{m_{p}^{n}}\operatorname{Im}\int_{C_{B/U,R}}\frac{p_{n}^{\pm
}\left(  r\right)  }{f^{\frac{n}{2}+1}\left(  r\right)  }dr,
\end{align}
where contributions from the ranges $\left(  r_{1},r_{h}-R\right)  $ and
$\left(  r_{h}+R,r_{2}\right)  $ are discarded since they are always real. If
the radii of the Laurent series of $\frac{p_{n}^{\pm}\left(  r\right)
}{f^{\frac{n}{2}+1}\left(  r\right)  }$ at $r=r_{h}$ are larger than $R$, we
can Laurent expand $\frac{p_{n}^{\pm}\left(  r\right)  }{f^{\frac{n}{2}%
+1}\left(  r\right)  }$ on $C_{B/U,R}$, and only the coefficients $a_{-1}$ of
$\left(  r-r_{h}\right)  ^{-1}$ terms contribute to the imaginary part of the
integrals. This justifies the procedure to obtain eqns. $\left(
\ref{eq:ImW}\right)  $ and $\left(  \ref{eq:ImW-massless}\right)  $. Since the
expansions for $p_{r}^{\pm}$ can be trusted on $C_{U/B,R}$, one has
$\frac{\omega}{\sqrt{f\left(  r\right)  }m_{p}}\sim\frac{\omega}{\sqrt{\kappa
R}m_{p}}\lesssim1$ on $C_{U/B,R}$, which gives $R\gtrsim\frac{\omega^{2}%
}{\kappa m_{p}^{2}}$. Usually, one has that the radii of the Laurent series of
$\frac{p_{n}^{\pm}\left(  r\right)  }{f^{\frac{n}{2}+1}\left(  r\right)  }%
\sim\kappa^{-1}$ and hence $\kappa^{-1}\gtrsim R$. For $R$ to exist, one has
$\kappa^{-1}\gtrsim\frac{\omega^{2}}{\kappa m_{p}^{2}}$ which leads to
$\omega\lesssim m_{p}$. Note that if Lorentz violating scale $\Lambda$ is much
smaller than $m_{p}$, one instead has that $\frac{\omega}{\sqrt{f\left(
r\right)  }\Lambda}\lesssim1$ on $C_{U/B,R}$ and $\omega\lesssim\Lambda$.

Various theories of quantum gravity, such as string theory, loop quantum
gravity and quantum geometry, predict the existence of a minimal length
\cite{TR-Townsend:1977xw,TR-Amati:1988tn,TR-Konishi:1989wk}. The generalized
uncertainty principle (GUP) \cite{TR-Kempf:1994su} is a simply way to realize
this minimal length. To incorporate the Klein-Gordon/Dirac equation with the
GUP, one usually considers the quantization in position representation. In
position representation, the operators $\vec{k}=-i\vec{\nabla}$ and
$\omega=i\partial_{t}$ are introduced \cite{TR-Hossenfelder:2003jz}. One then
can express the energy and momentum operators as functions of $\vec{k}$ and
$\omega$ and obtain the deformed Klein-Gordon/Dirac equations in flat
spacetime. Inserting the ansatz $\varphi=\exp\left(  iEt-i\vec{p}\cdot\vec
{x}\right)  $ in the Klein-Gordon/Dirac equation gives the dispersion relation
for $E$ and $p$ in flat spacetime. In
\cite{IN-Chen:2013pra,IN-Chen:2013tha,IN-Chen:2013ssa,IN-Chen:2014xsa,IN-Chen:2014xgj}%
, the deformed Dirac equation was generalized to curved spacetime. The
modified Hawking temperatures of various black holes were then derived via the
Hamilton-Jacobi method. In our appendix, the Dirac equation is generalized to
curved spacetime for any preferred frame. In particular, the static frame is
used in our paper. It turns out that the way of generalizing the Dirac
equation to curved spacetime in
\cite{IN-Chen:2013pra,IN-Chen:2013tha,IN-Chen:2013ssa,IN-Chen:2014xsa,IN-Chen:2014xgj}
is the same as that in our paper. Thus, we can use the dispersion relation for
$E$ and $p$ obtained in flat spacetime and eqn. $\left(  \ref{eq:delta}%
\right)  $ to reproduce the modified Hawking temperatures of black holes with
the metric $\left(  \ref{eq:BHmetric}\right)  $ obtained in
\cite{IN-Chen:2013pra,IN-Chen:2013tha,IN-Chen:2013ssa,IN-Chen:2014xsa,IN-Chen:2014xgj}%
. In fact, the dispersion relation in flat spacetime in
\cite{IN-Chen:2013pra,IN-Chen:2013tha,IN-Chen:2013ssa,IN-Chen:2014xsa,IN-Chen:2014xgj}
is given by
\begin{equation}
p^{2}\approx E^{2}-m^{2}+2\tilde{\beta}E^{4}+\mathcal{O}\left(  \tilde{\beta
}^{2}\right)  ,
\end{equation}
which by comparing to eqn. $\left(  \ref{eq:MDRExpansion}\right)  $ gives
\begin{equation}
\alpha=1,\beta=1,\gamma=0,C_{3}=0,C_{4}=2,\text{ and }m_{p}=\frac{1}%
{\sqrt{\tilde{\beta}}}\text{.}%
\end{equation}
Thus, eqn. $\left(  \ref{eq:delta}\right)  $ becomes%
\begin{equation}
\Delta_{qm}=\tilde{\beta}\left[  \frac{24\tilde{\omega}^{\prime}\left(
r_{h}\right)  \tilde{\omega}\left(  r_{h}\right)  }{16\kappa}+\frac{1}%
{2}\left(  \frac{\lambda}{C\left(  r_{h}^{2}\right)  }+m^{2}\right)
-\frac{f^{\prime\prime}\left(  r_{h}\right)  \tilde{\omega}^{2}\left(
r_{h}\right)  }{4\kappa^{2}}\right]  . \label{eq:delta-GUP}%
\end{equation}
For example, $\Delta_{qm}$ for a particle with the angular momentum $l=0$ in a
Schwarzschild black hole with the mass $M$
\cite{IN-Chen:2013pra,IN-Chen:2014xgj} is%
\begin{equation}
\Delta_{qm}=\frac{\tilde{\beta}}{2}\left(  m^{2}+4\omega^{2}\right)  .
\end{equation}
$\Delta_{qm}$ for a neutral particle with the angular momentum $l=0$ in a
Reissner-Nordstrom with the mass $M$ and the charge $Q$
\cite{IN-Chen:2013tha,IN-Chen:2014xgj} is%
\begin{equation}
\Delta_{qm}=\tilde{\beta}\left[  \frac{m^{2}}{2}+\frac{2r_{+}\left(
r_{+}-2r_{-}\right)  }{\left(  r_{+}-r_{-}\right)  ^{2}}\omega^{2}\right]  ,
\end{equation}
where $r_{\pm}=M\pm\sqrt{M^{2}-Q^{2}}$. Here we only consider a neutral
particle to make a comparison since the electromagnetic filed was included in
\cite{IN-Chen:2013tha,IN-Chen:2014xgj} in a different way.

Following the argument proposed in \cite{IN-AmelinoCamelia:2004xx}, the
authors in \cite{IN-AmelinoCamelia:2005ik} obtained modified relations between
the mass of a Schwarzschild black hole and its entropy and temperature. The
argument connecting a MDR and some modifications of the entropy of black holes
is formulated in a scheme of analysis first introduced by Bekenstein
\cite{IN-Bekenstein:1973ur}. In fact, the modified temperature of the black
hole for the MDR $\left(  \ref{eq:MDRExpansion}\right)  $ with $m=0$ was given
by
\begin{equation}
T=T_{0}\left[  1-\frac{m_{p}C_{3}}{4M}+\frac{m_{p}^{2}}{4M^{2}}\left(
\frac{5C_{3}^{2}}{8}-\frac{C_{4}}{2}\right)  +\mathcal{O}\left(  \frac
{m_{p}^{3}}{M^{3}}\right)  \right]  ,\label{eq:Temp-MDR}%
\end{equation}
where $M$ is the mass of the black hole, and $T_{0}=\frac{m_{p}^{2}}{8\pi M}$.
Note that eqn. $\left(  \ref{eq:Temp-MDR}\right)  $ is obtained by using eqn.
$\left(  16\right)  $ of \cite{IN-AmelinoCamelia:2005ik}. Eqn. $\left(
16\right)  $ of \cite{IN-AmelinoCamelia:2005ik} gave the temperature of the
black hole
\begin{equation}
T=\frac{1}{4\pi}f_{disp}\left(  \frac{m_{p}^{2}}{2M}\right)  ,
\end{equation}
where $E=$ $f_{disp}\left(  p\right)  $, and $f_{disp}\left(  p\right)  $ for
the MDR $\left(  \ref{eq:MDRExpansion}\right)  $ with $m=0$ is%
\begin{equation}
f_{disp}\left(  p\right)  \approx p\left[  1-\frac{C_{3}p}{2m_{p}}+\left(
\frac{5C_{3}^{2}}{8}-\frac{C_{4}}{2}\right)  \frac{p^{2}}{m_{p}^{2}}\right]  .
\end{equation}
On the other hand, we can use eqn. $\left(  \ref{eq:delta}\right)  $ to
estimate the temperature of the black hole. For a massless particle in a
Schwarzschild black hole, eqn. $\left(  \ref{eq:delta}\right)  $ gives%
\begin{equation}
\Delta=\frac{1}{32m_{p}^{2}}\left[  \left(  4C_{4}-3C_{3}^{2}\right)
\frac{\lambda}{2M^{2}}+8\left(  4C_{4}-C_{3}^{2}\right)  \omega^{2}\right]
+\mathcal{O}\left(  m_{p}^{-3}\right)  ,
\end{equation}
where $\lambda$ is the magnitude of the angular momentum of the particle. The
event horizon of the Schwarzschild black hole is $r_{h}=2M$. Near the horizon
of the the black hole, one has $\lambda\sim\left(  pr_{h}\right)  ^{2}%
\sim\left(  \omega r_{h}\right)  ^{2}$. Thus, one can rewrite $\Delta$%
\begin{equation}
\Delta\sim\frac{\omega^{2}}{16m_{p}^{2}}\left(  20C_{4}-7C_{3}^{2}\right)
+\mathcal{O}\left(  m_{p}^{-3}\right)  .
\end{equation}
As reported in \cite{IN-AmelinoCamelia:2005ik,IN-Nozari:2006ka,IN-Ling:2005bq}%
, the authors obtained the relation $\omega\gtrsim\frac{\hbar}{\delta
x}+\mathcal{O}\left(  \frac{1}{m_{p}}\right)  $ between the energy of a
particle and its position uncertainty for a MDR. Near the horizon of the
Schwarzschild black hole, the position uncertainty of a particle is of the
order of the Schwarzschild radius of the black hole
\cite{IN-Bekenstein:1973ur} $\delta x\sim r_{h}=2M$. Thus, one finds for $T$
that%
\begin{equation}
T\sim T_{0}\left[  1+\frac{m_{p}^{2}}{64M^{2}}\left(  7C_{3}^{2}%
-20C_{4}\right)  +\mathcal{O}\left(  \frac{m_{p}^{3}}{M^{3}}\right)  \right]
.\label{eq:Temp-HJ}%
\end{equation}
From eqns. $\left(  \ref{eq:Temp-MDR}\right)  $ and $\left(  \ref{eq:Temp-HJ}%
\right)  $, it indicates that the heuristics methods used in
\cite{IN-AmelinoCamelia:2005ik} and our paper give different estimations of
the black hole's temperature. Strictly speaking, in our model particles with
different energy $\omega$ and angular momentum $\lambda$ would have different
effective Hawking temperature so that, in general, a \textquotedblleft
unique\textquotedblright\ equilibrium temperature is not well defined.
However, we assume an average behavior for any particle described by a unique
average temperature for the system. This average temperature for massless
particles around a Schwarzschild black hole is given in eqn. $\left(
\ref{eq:Temp-HJ}\right)  $, which is obtained from the effective Hawking
temperature $\left(  \ref{eq:TeffMassless}\right)  $ by estimating
$\lambda\sim\left(  \omega r_{h}\right)  ^{2}$ and $\omega\sim\hbar/r_{h}$. On
the other hand, the authors of \cite{IN-AmelinoCamelia:2005ik} first followed
the original Bekenstein argument to find the  modified relation between the
mass of a Schwarzschild black hole and its entropy due to MDR. Then, the
modified black hole temperature $\left(  \ref{eq:Temp-MDR}\right)  $ was
obtained by using the first law of black hole thermodynamics. 

\section{Thermodynamics of Radiations}

\label{Sec:TOR}

For particles emitted in a wave mode labelled by energy $\omega$ and $\lambda$
plus some other quantum numbers $J_{i}$ if needed, we find that%
\begin{align*}
&  \left(  \text{Probability for a black hole to emit a particle in this
mode}\right) \\
&  =\exp\left(  -\frac{\omega}{T_{eff}}\right)  \times(\text{Probability for a
black hole to absorb a particle in the same mode}),
\end{align*}
where $T_{eff}$ is given by eqns. $\left(  \ref{eq:TeffMassive}\right)  $ or
$\left(  \ref{eq:TeffMassless}\right)  $. The above relation for usual
dispersion relation was obtained by Hartle and Hawking
\cite{TOR-Hartle:1976tp} using semiclassical analysis. If the black hole is in
equilibrium, the rate of emission particles by the black hole must exactly
equal the rate of absorption. Neglecting back-reaction, detailed balance
condition requires that the ratio of the probability of having $N$ particles
in a particular mode with $\omega,\lambda$ and $J_{i}$ to the probability of
having $N-1$ particles in the same mode is $\exp\left(  -\frac{\omega}%
{T_{eff}}\right)  .$ Thus, we find that the probability of having $N$
particles $P_{N}\left(  \omega,\lambda,J_{i}\right)  $ in the mode is given by%
\begin{equation}
P_{N}\left(  \omega,\lambda,J_{i}\right)  =C_{\omega,\lambda,J_{i}}\exp\left(
-\frac{N\omega}{T_{eff}}\right)  ,
\end{equation}
where $C_{\omega,\lambda,J_{i}}$ is a normalizing constant. $C_{\omega
,\lambda,J_{i}}$ is determined by the normalized condition $%
{\displaystyle\sum\limits_{N=0}^{N_{\infty}}}
P_{N}\left(  \omega,\lambda,J_{i}\right)  =1$ where $N_{\infty}=\infty$ for
bosons and $N_{\infty}=1$ for fermions. Thus, the probability $P_{N}\left(
\omega,\lambda,J_{i}\right)  $ is%
\begin{equation}
P_{N}\left(  \omega,\lambda,J_{i}\right)  =\left[  1-\left(  -1\right)
^{\epsilon}\exp\left(  -\frac{\omega}{T_{eff}}\right)  \right]  ^{1-2\epsilon
}\exp\left(  -\frac{N\omega}{T_{eff}}\right)  , \label{eq:ProbN}%
\end{equation}
where $\epsilon=0$ for bosons and $\epsilon=1$ for fermions. To calculate the
average number $n_{\omega,\lambda,J_{i}}$ in the mode, we define
\begin{equation}
A_{\omega,\lambda,J_{i}}\left(  \mu\right)  =%
{\displaystyle\sum\limits_{N=0}^{N_{\infty}}}
\exp\left(  N\mu-\frac{N\omega}{T_{eff}}\right)  ,
\end{equation}
where one has $C_{\omega,\lambda,J_{i}}=A_{\omega,\lambda,J_{i}}^{-1}\left(
0\right)  $. So we find
\begin{equation}
n_{\omega,\lambda,J_{i}}=%
{\displaystyle\sum\limits_{N=0}^{N_{\infty}}}
NP_{N}\left(  \omega,\lambda,J_{i}\right)  =\frac{\partial_{\mu}%
A_{\omega,\lambda,J_{i}}\left(  \mu\right)  }{A_{\omega,\lambda,J_{i}}\left(
\mu\right)  }|_{\mu=0}=\frac{1}{\exp\left(  \frac{\omega}{T_{eff}}\right)
-\left(  -1\right)  ^{\epsilon}}. \label{eq:AverageNumber}%
\end{equation}
Using eqns. $\left(  \ref{eq:ProbN}\right)  $ and $\left(
\ref{eq:AverageNumber}\right)  $, one can rewrite $P_{N}\left(  \omega
,\lambda,J_{i}\right)  $ in terms of $n_{\omega,\lambda,J_{i}}$ as%
\begin{equation}
P_{N}\left(  \omega,\lambda,J_{i}\right)  =n_{\omega,\lambda,J_{i}}^{N}\left[
1+\left(  -1\right)  ^{\epsilon}n_{\omega,\lambda,J_{i}}\right]  ^{-N-\left(
-1\right)  ^{\epsilon}},
\end{equation}
where $N$ can be any non-negative integer for bosons ($\epsilon=0$) but is
restricted to be $0$ or $1$ for fermions ($\epsilon=1$). The von Neumann
entropy for the mode is%
\begin{align}
s_{\omega,\lambda,J_{i}}  &  =-%
{\displaystyle\sum\limits_{N=0}^{N_{\infty}}}
P_{N}\left(  \omega,\lambda,J_{i}\right)  \ln P_{N}\left(  \omega
,\lambda,J_{i}\right)  ,\nonumber\\
&  =\left[  n_{\omega,\lambda,J_{i}}+\left(  -1\right)  ^{\epsilon}\right]
\ln\left[  1+\left(  -1\right)  ^{\epsilon}n_{\omega,\lambda,J_{i}}\right]
-n_{\omega,\lambda,J_{i}}\ln n_{\omega,\lambda,J_{i}}
\label{eq:EntropyForMode}%
\end{align}
where we use $%
{\displaystyle\sum\limits_{N=0}^{N_{\infty}}}
NP_{N}\left(  \omega,\lambda,J_{i}\right)  =n_{\omega,\lambda,J_{i}}$. The
total entropy of radiation is
\begin{equation}
S=\sum\limits_{\omega,\lambda,J_{i}}s_{\omega,\lambda,J_{i}}\text{,}%
\end{equation}
which will be calculated in the brick wall model in section \ref{Sec:EBW}.
Note that since $T_{eff}$ only depends on $\omega$ and $\lambda$, the average
number $n_{\omega,\lambda,J_{i}}$ and the entropy $s_{\omega,\lambda,J_{i}}$
are independent of $J_{i}$. Thus, we could omit the subscript $J_{i}$ in
$n_{\omega,\lambda,J_{i}}$ and $s_{\omega,\lambda,J_{i}}$ from now on.
Defining $n\left(  x\right)  $ and $s\left(  x\right)  $ by
\begin{align}
n\left(  x\right)   &  =\frac{1}{\exp x-\left(  -1\right)  ^{\epsilon}%
},\nonumber\\
s\left(  x\right)   &  =\frac{\left(  -1\right)  ^{\epsilon}\exp x}{\exp
x-\left(  -1\right)  ^{\epsilon}}\ln\left[  \frac{\exp x}{\exp x-\left(
-1\right)  ^{\epsilon}}\right]  +\frac{\ln\left[  \exp x-\left(  -1\right)
^{\epsilon}\right]  }{\exp x-\left(  -1\right)  ^{\epsilon}},
\end{align}
we can write $n_{\omega,\lambda}$ and $s_{\omega,\lambda}$ with respect to
$n\left(  x\right)  $ and $s\left(  x\right)  $%
\begin{align}
n_{\omega,\lambda}  &  =n\left(  \frac{\omega}{T_{eff}}\right)  ,\nonumber\\
s_{\omega,\lambda}  &  =s\left(  \frac{\omega}{T_{eff}}\right)  \text{.}
\label{eq:Entropy}%
\end{align}

When integrating over $\omega$, we need to specify the upper limit on the
energy of the emitted particle. One of the limits comes from the requirement
that the energy of the particle could not exceed the mass of the black hole.
Another one is from the effective field theories in the appendix. Suppose that
the higher dimensional operators in the effective field theories are
suppressed by some scale of Lorentz violation $\Lambda$. Usually, we can only
trust the effective theories below $\Lambda$. As decoupling theorem
\cite{TOR-Appelquist:1974tg} shows, the contributions above $\Lambda$ in some
regularized theory gets absorbed into Wilson coefficients of the effective
theories, $Cs$ and $Bs$ in the appendix. Consequently, the energy of the
particle could not exceed $\Lambda$ otherwise our effective theories would
break down. Note that $\omega\lesssim\Lambda$ has also been obtained in
section \ref{Sec:TR}. Thus, the energy of the emitted particle $\omega
\leq\omega_{\max}\equiv\min\left\{  M,\Lambda\right\}  $. In the remaining of
the paper, we would encounter the integrals like%
\begin{equation}
\int_{0}^{u_{\max}}u^{i}n\left(  u\right)  du\text{ or }\int_{0}^{u_{\max}%
}u^{i}s\left(  u\right)  du, \label{eq:N&SIntegral}%
\end{equation}
where $u_{\max}=\frac{\omega_{\max}}{T_{0}}$ and $i$ is a non-negative
integer. For a black hole with the mass $M\gg m_{p}$, one finds that $u_{\max
}=\frac{2\pi m_{p}}{m_{p}^{2}\kappa}\sim\frac{1}{\kappa m_{p}}\gg1$. For
example, $\kappa=\frac{1}{4M\text{ }}$ and $\kappa m_{p}\ll1$ in the
Schwarzschild metric. For such case, using $n\left(  x\right)  \sim e^{-x}$
and $s\left(  x\right)  \sim xe^{-x}$ for $x\gg1$, one gets%
\begin{equation}
\int_{u_{\max}}^{\infty}u^{i}n\left(  u\right)  du\sim\frac{e^{-\frac
{1}{\kappa m_{p}}}}{\left(  \kappa m_{p}\right)  ^{i}}\text{ and }%
\int_{u_{\max}}^{\infty}u^{i}s\left(  u\right)  du\sim\frac{e^{-\frac
{1}{\kappa m_{p}}}}{\left(  \kappa m_{p}\right)  ^{i+1}},
\end{equation}
which can be safely neglected for $\kappa m_{p}\ll1$, and hence we can let
$u_{\max}=\infty$ in eqn. $\left(  \ref{eq:N&SIntegral}\right)  $. Therefore,
in section \ref{Sec:EBW}, we let $u_{\max}=\infty$ for integrals of the form
in eqn. $\left(  \ref{eq:N&SIntegral}\right)  $ since we are only interested
in the divergent part of the entanglement entropy as $\kappa m_{p}%
\rightarrow0$.

\section{Entropy in Brick Wall Model}

\label{Sec:EBW}

In 1985 t' Hooft \cite{EBW-'tHooft:1984re}\ proposed the brick wall model to
calculate the entropy of a thermal gas of Hawking particles propagating just
outside the black hole horizon. The entropy is calculated by methods of the
WKB approximation. However, when it comes to calculate the density of states
of emitted particles, t' Hooft found that they became infinite as one got
closer to the horizon. To make the entropy finite, he introduced a brick wall
cut-off near the horizon such that the boundary condition%
\begin{equation}
\Phi\left(  x\right)  =0\text{ \ at }r=r_{h}+r_{\varepsilon},
\end{equation}
where $\Phi$ is the radiation's field. Moreover, another cut-off at a large
distance from the horizon $L\gg r_{h}$ was introduced to eliminate infrared divergences.

For simplicity, we consider in this section\ a 4D spherically symmetric black
hole with the metric%
\begin{equation}
ds^{2}=f\left(  r\right)  dt^{2}-\frac{1}{f\left(  r\right)  }dr^{2}-C\left(
r^{2}\right)  d\Omega.
\end{equation}
For such a black hole, the quantum numbers needed to specify a wave mode of
radiation are the energy $\omega$, the angular momentum $l$, the magnetic
quantum number $m$. We also assume that the radiated particles are massless
and neutral. Thus, the MDR $\left(  \ref{eq:ExactMDR}\right)  $, becomes%
\begin{equation}
p^{2}=m_{p}^{2}H\left(  \frac{E}{m_{p}},0\right)  \equiv m_{p}^{2}H\left(
\frac{E}{m_{p}}\right)  , \label{eq:MDRexactMassless}%
\end{equation}
where we define $H\left(  x\right)  =H\left(  x,0\right)  $, and $H\left(
x\right)  \sim x^{2}$ for $x\ll1$. As shown in section \ref{Sec:DHJE}, the
deformed Hamilton-Jacobi equation incorporating the MDR $\left(
\ref{eq:MDRexactMassless}\right)  $ for a massless and neutral particle in the
4D spherically symmetric black hole is given by%
\begin{equation}
X^{2}=m_{p}^{2}H\left(  \frac{T}{m_{p}}\right)  ,
\label{eq:deformedHJMassless}%
\end{equation}
where
\begin{equation}
T=\frac{\omega}{\sqrt{f\left(  r\right)  }},X^{2}=f\left(  r\right)  p_{r}%
^{2}+\frac{\left(  l+\frac{1}{2}\right)  ^{2}\hbar^{2}}{C\left(  r^{2}\right)
}.
\end{equation}
Then, we get from eqn. $\left(  \ref{eq:deformedHJMassless}\right)  $ that \
\begin{equation}
p_{r}^{2}=\frac{1}{f\left(  r\right)  }\left[  m_{p}^{2}H\left(  \frac{\omega
}{m_{p}\sqrt{f\left(  r\right)  }}\right)  -\frac{\left(  l+\frac{1}%
{2}\right)  ^{2}\hbar^{2}}{C\left(  r^{2}\right)  }\right]  . \label{eq:Pr}%
\end{equation}
Define the radial wave number $k\left(  r,l,\omega\right)  $ by%
\begin{equation}
k\left(  r,l,\omega\right)  =\left\vert p_{r}\right\vert ,
\end{equation}
as long as $p_{r}^{2}\geq0$, and $k\left(  r,l,\omega\right)  =0$ otherwise.
Taking two Dirichlet conditions at $r=r_{h}+r_{\varepsilon}$ and $r=L$ into
account, one finds that the number of one-particle states not exceeding
$\omega$ with fixed value of the angular momentum $l$ is given by%
\begin{equation}
n\left(  \omega,l\right)  =\frac{1}{\pi\hbar}\int_{r_{h}+r_{\varepsilon}}%
^{L}k\left(  r,l,\omega\right)  dr.
\end{equation}
Thus, we obtain for the total entropy of radiation that
\begin{align}
S  &  =\sum\limits_{\omega,l,m}s_{\omega,l}=\int\left(  2l+1\right)  dl\int
d\omega\frac{dn\left(  \omega,l\right)  }{d\omega}s_{\omega,l}\nonumber\\
&  =\frac{1}{\pi\hbar}\int\left(  2l+1\right)  dl\int d\omega\int
_{r_{h}+r_{\varepsilon}}^{L}dr\frac{dk\left(  r,l,\omega\right)  }{d\omega
}s_{\omega,l}\nonumber\\
&  =\frac{C\left(  r_{h}^{2}\right)  m_{p}^{2}}{2\pi\hbar^{3}}\int
_{r_{h}+r_{\varepsilon}}^{L}\frac{dr}{f\left(  r\right)  }\int d\omega
H^{\prime}\left(  \frac{\omega}{m\sqrt{f\left(  r\right)  }}\right)
\nonumber\\
\text{ }  &  \text{\ \ }\int_{0}^{H\left(  \frac{\omega}{m_{p}\sqrt{f\left(
r\right)  }}\right)  \frac{C\left(  r^{2}\right)  }{C\left(  r_{h}^{2}\right)
}}dz\left[  H\left(  \frac{\omega}{m_{p}\sqrt{f\left(  r\right)  }}\right)
-z\frac{C\left(  r_{h}^{2}\right)  }{C\left(  r^{2}\right)  }\right]
^{-\frac{1}{2}}s_{\omega,l}, \label{eq:EntropySWL}%
\end{align}
where we define a dimensionless parameter $z=\frac{\left(  l+\frac{1}%
{2}\right)  ^{2}\hbar^{2}}{C\left(  r_{h}^{2}\right)  m_{p}^{2}}$. Using
$\lambda=\left(  l+\frac{1}{2}\right)  ^{2}\hbar^{2}$ and $z=\frac{\left(
l+\frac{1}{2}\right)  ^{2}\hbar^{2}}{C\left(  r_{h}^{2}\right)  m_{p}^{2}}$,
one rewrites eqn. $\left(  \ref{eq:DeltaMassless}\right)  $ as%
\begin{equation}
\Delta=%
{\displaystyle\sum\limits_{a=0}^{\infty}}
{\displaystyle\sum\limits_{k=0}^{\infty}}
\eta_{a}^{2k+2a}\zeta_{k}^{a}z^{a}\frac{\omega^{2k}}{m_{p}^{2k}}-1.
\end{equation}
Defining the coefficients $\xi_{l,k}^{n}$ by%
\begin{equation}
\left(
{\displaystyle\sum\limits_{l^{\prime}=0}^{\infty}}
{\displaystyle\sum\limits_{k^{\prime}=0}^{\infty}}
\eta_{l^{\prime}}^{2k^{\prime}+2l^{\prime}}\zeta_{k^{\prime}}^{l^{\prime}%
}z^{l^{\prime}}x^{2k^{\prime}}-1\right)  ^{n}=%
{\displaystyle\sum\limits_{k=0}^{\infty}}
{\displaystyle\sum\limits_{l=0}^{\infty}}
\xi_{l,k}^{n}z^{l}x^{2k}, \label{eq:Xsi}%
\end{equation}
one has for $s_{\omega,l}$%
\begin{align}
s_{\omega,l}  &  =s\left(  \frac{\omega}{T_{eff}}\right)  =s\left(
\frac{\omega\left(  1+\Delta\right)  }{T_{0}}\right) \label{eq:EntropyZ}\\
&  =\sum_{n=0}\frac{1}{n!}s^{\left(  n\right)  }\left(  \frac{\omega}{T_{0}%
}\right)  \frac{\omega^{n}\Delta^{n}}{T_{0}^{n}}\nonumber\\
&  =%
{\displaystyle\sum\limits_{a=0}^{\infty}}
\sum_{n=0}^{\infty}%
{\displaystyle\sum\limits_{k=0}^{\infty}}
\left[  \frac{s^{\left(  n\right)  }\left(  u\right)  u^{n}}{n!}\right]
\xi_{a,k}^{n}z^{a}\left(  \frac{T_{0}u}{m_{p}}\right)  ^{2k}\nonumber\\
&  =\sum_{a=0}^{\infty}z^{a}\Theta_{a}\left(  u\right)  ,
\end{align}
where we define $u=\frac{\omega}{T_{0}}$ and%
\begin{equation}
\Theta_{a}\left(  u\right)  =%
{\displaystyle\sum\limits_{k=0}^{\infty}}
\left(  \frac{T_{0}u}{m_{p}}\right)  ^{2k}%
{\displaystyle\sum\limits_{n=0}^{k+a}}
\left[  \frac{s^{\left(  n\right)  }\left(  u\right)  u^{n}}{n!}\right]
\xi_{a,k}^{n}\text{.} \label{eq:Theta}%
\end{equation}
Note that $\xi_{k,l}^{0}=0$ except $\xi_{0,0}^{0}=1,\xi_{l,k<n-l}^{n}=0$ and
$\xi_{l<n-k,k}^{n}=0$. Putting eqn. $\left(  \ref{eq:EntropyZ}\right)  $ into
eqn. $\left(  \ref{eq:EntropySWL}\right)  $ and integrating eqn. $\left(
\ref{eq:EntropySWL}\right)  $ over $z$ gives%
\begin{align}
S  &  =\frac{C\left(  r_{h}^{2}\right)  m_{p}^{2}T_{0}}{2\pi\hbar^{3}}%
\sum_{a=0}^{\infty}\frac{\sqrt{\pi}a!}{\Gamma\left(  a+\frac{3}{2}\right)
}\int_{0}^{\infty}\Theta_{a}\left(  u\right)  du\nonumber\\
&  \text{ \ }\int_{r_{h}+r_{\varepsilon}}^{L}\frac{dr}{f\left(  r\right)
}\frac{C^{a+1}\left(  r^{2}\right)  }{C^{a+1}\left(  r_{h}^{2}\right)
}H^{a+\frac{1}{2}}\left(  \frac{T_{0}u}{m_{p}\sqrt{f\left(  r\right)  }%
}\right)  H^{\prime}\left(  \frac{T_{0}u}{m_{p}\sqrt{f\left(  r\right)  }%
}\right)  . \label{eq:EntopyAU}%
\end{align}
To calculate $S$, the variable $r$ may be changed by introducing
$x=\frac{T_{0}u}{m_{p}\sqrt{f\left(  r\right)  }}$. Then, eqn. $\left(
\ref{eq:EntopyAU}\right)  $ becomes%
\begin{equation}
S=\frac{C\left(  r_{h}^{2}\right)  m_{p}^{2}}{4\pi^{2}\hbar^{2}}\sum
_{a=0}^{\infty}\frac{\sqrt{\pi}a!}{\Gamma\left(  a+\frac{3}{2}\right)  }%
\int_{0}^{\infty}\Theta_{a}\left(  u\right)  du\int_{\delta}^{x_{\varepsilon}%
}x^{-1}H^{a+\frac{1}{2}}\left(  x\right)  H^{\prime}\left(  x\right)
G_{a}\left(  \frac{u^{2}T_{0}^{2}}{x^{2}m_{p}^{2}}\right)  dx,
\label{eq:EntropyGx}%
\end{equation}
where we define $x_{\varepsilon}=\frac{T_{0}u}{m_{p}\sqrt{f\left(
r_{h}+r_{\varepsilon}\right)  }}$, $\delta=\frac{T_{0}u}{m_{p}}\frac{1}%
{\sqrt{f\left(  L\right)  }}$, and%
\begin{equation}
G_{a}\left(  y\right)  =\frac{2\kappa C^{a+1}\left[  f^{-1}\left(  y\right)
^{2}\right]  }{C^{a+1}\left(  r_{h}^{2}\right)  f^{\prime}\left[
f^{-1}\left(  y\right)  \right]  }\text{.}%
\end{equation}
Now the brick walls are at $x=x_{\varepsilon}$ and $x=\delta$. Note that
$x_{\varepsilon}\rightarrow\infty$ when $r_{\varepsilon}\rightarrow0$, and the
horizon is at $x_{\varepsilon}=\infty$. Since $G_{a}\left(  0\right)  =1$, we
can Taylor expand $G_{a}\left(  y\right)  $ at $y=0$%
\begin{equation}
G_{a}\left(  y\right)  =%
{\displaystyle\sum\limits_{k=0}^{\infty}}
f_{k}^{a}y^{k}, \label{eq:G-Expansion}%
\end{equation}
where we find the first two coefficients of the series expansion are
$f_{0}^{a}=1$ and $f_{1}^{a}=\frac{r_{h}C^{\prime}\left(  r_{h}^{2}\right)
}{\kappa C\left(  r_{h}^{2}\right)  }\left[  \left(  a+1\right)
-\frac{C\left(  r_{h}^{2}\right)  f^{\prime\prime}\left(  r_{h}\right)
}{4\kappa r_{h}C^{\prime}\left(  r_{h}^{2}\right)  }\right]  $. Substituting
eqn. $\left(  \ref{eq:G-Expansion}\right)  $ into eqn. $\left(
\ref{eq:EntropyGx}\right)  $ gives us%
\begin{equation}
S=\frac{C\left(  r_{h}^{2}\right)  }{4\pi^{2}m_{p}^{2}}%
{\displaystyle\sum\limits_{k=0}^{\infty}}
\left(  \frac{m_{p}\kappa}{2\pi}\right)  ^{2k}\sum_{a=0}^{\infty}\frac
{\sqrt{\pi}a!f_{k}^{a}}{\Gamma\left(  a+\frac{3}{2}\right)  }\int_{0}^{\infty
}u^{2k}\Theta_{a}\left(  u\right)  du\int_{\delta}^{x_{\varepsilon}}%
H^{a+\frac{1}{2}}\left(  x\right)  H^{\prime}\left(  x\right)  x^{-2k-1}dx,
\label{eq:EntropyHx}%
\end{equation}
where we use $T_{0}=\frac{\hbar\kappa}{2\pi}$ and $\hbar=$ $m_{p}^{2}$. The
entropy receives two contributions, one from the horizon and the other from
the vacuum surrounding the system at large distances. The second one is
irrelevant for our purposes and henceforth discarded.

For the usual scenario with $H\left(  x\right)  =x^{2}$, the integrals over
$x$ in eqn. $\left(  \ref{eq:EntropyHx}\right)  $ become divergent for $a=k=0$
as one approaches the horizon with $x_{\varepsilon}\rightarrow\infty$. This
divergence leads to the introduction of the wall near the horizon by t' Hooft.
However, the $x$-integrals could be finite as $x_{\varepsilon}\rightarrow
\infty$ for some MDRs. In fact, there are two kinds of MDRs for the integrals
to be finite. For the first kind of these MDRs, the high energy contributions
are suppressed. For example, the \textquotedblleft all-order
MDR\textquotedblright\ of form%
\begin{equation}
p^{2}=2m_{p}^{2}\exp\left(  -\frac{E}{m_{p}}\right)  \left[  \cosh\left(
\frac{E}{m_{p}}\right)  -1\right]  , \label{eq:Kappa-MDR}%
\end{equation}
was given in the $\kappa$-Minkowski noncommutative spacetime in
\cite{EBW-KowalskiGlikman:2002jr}. For such a MDR, one has%
\begin{gather}
H\left(  x\right)  =2\exp\left(  -x\right)  \left[  \cosh\left(  x\right)
-1\right]  ,\nonumber\\
H^{l+\frac{1}{2}}\left(  x\right)  H^{\prime}\left(  x\right)  x^{-2k-1}%
\sim\exp\left(  -x\right)  x^{-2k-1}\text{ as }x\rightarrow\infty,
\end{gather}
which guaranties the convergence of the $x$-integrals as $x_{\varepsilon
}\rightarrow\infty$. Another example is inspired by the all order generalized
uncertainty relation considered in \cite{IN-Kim:2007if}. The MDR can be
written as%
\begin{equation}
\frac{dp}{dE}=\exp\left(  -\frac{E^{2}}{m_{p}^{2}}\right)  ,
\end{equation}
which gives
\begin{gather}
H\left(  x\right)  =\left(  \int_{0}^{x}e^{-x^{2}}dx\right)  ^{2},\nonumber\\
H^{l+\frac{1}{2}}\left(  x\right)  H^{\prime}\left(  x\right)  x^{-2k-1}%
\lesssim\left(  e^{-x^{2}}\right)  ^{2l+3}x^{-2k-1}\text{ as }x\rightarrow
\infty\text{.}%
\end{gather}
Thus, the $x$-integrals stay finite as $x\rightarrow\infty.$ Moreover, it was
found in \cite{IN-Kim:2007if} that the entropy kept finite when the wall
approached the horizon, and hence the wall in the brick wall model located
just outside the horizon could be avoided. For the second kind of the MDRs,
the energy $E$ in the MDRs has a maximum value, and hence $x_{\varepsilon}$
can not go to the infinity. For example, Corley and
Jacobson\cite{IN-Corley:1996ar} proposed%
\begin{equation}
E=\sqrt{p^{2}-\frac{p^{4}}{4m_{p}^{2}}},
\end{equation}
which gives%
\begin{align}
H\left(  x\right)   &  =2\left(  1-\sqrt{1-x}\right)  \text{ for }0\leq
x\leq1,\nonumber\\
H^{l+\frac{1}{2}}\left(  x\right)  H^{\prime}\left(  x\right)  x^{-2k-1}  &
\sim\frac{1}{\sqrt{1-x}}\text{ as }x\rightarrow1.
\end{align}
So the $x$-integrals are finite for the Corley and Jacobson dispersion
relation. It was shown in \cite{IN-Chang:2003sa} that the entropy was rendered
UV finite for the Corley and Jacobson dispersion relation. For the Unruh
dispersion relation \cite{IN-Unruh:1994je}%
\begin{equation}
E=m_{p}\left[  \tanh\left(  \frac{p^{n}}{m_{p}^{n}}\right)  \right]
^{\frac{1}{n}},
\end{equation}
we have%
\begin{align}
H\left(  x\right)   &  =\left[  \tanh^{-1}\left(  x^{n}\right)  \right]
^{\frac{1}{n}}\text{ for }0\leq x\leq1,\nonumber\\
H^{l+\frac{1}{2}}\left(  x\right)  H^{\prime}\left(  x\right)  x^{-2k-1}  &
\sim\frac{\left[  \ln\left(  1-x^{n}\right)  \right]  ^{\left(  l+\frac{3}%
{2}\right)  \frac{1}{n}-1}}{1-x^{n}}\text{ as }x\rightarrow1.
\end{align}
We find that the $x$-integrals diverge as $x\rightarrow1$, and a wall near the
horizon is needed. However, the entropy for the Unruh dispersion relation was
also found finite in \cite{IN-Chang:2003sa}. This might be due to different
modifications of the dispersion relation in \cite{IN-Chang:2003sa} and our
paper. In particular, from eqn. $\left(  16\right)  $ in
\cite{IN-Chang:2003sa}, it showed that only the MDR along the radial direction
was modified by the transplanckian effect in \cite{IN-Chang:2003sa}. In the
remaining of section, we will consider two cases, in one of which the
$x$-integrals converge, and in the other they diverge.

\subsection{UV Finite Case}

We here assume that the $x$-integrals converge as $x\rightarrow\Lambda$, where
$\Lambda=\infty$ for the first kind of the MDRs in this case and $\Lambda$ is
the largest $x$ for the second kind. Thus, we can define%
\begin{equation}
\tilde{c}_{k}^{a}=\int_{\delta}^{\Lambda}H^{a+\frac{1}{2}}\left(  x\right)
H^{\prime}\left(  x\right)  x^{-2k-1}dx.
\end{equation}
Since $H\left(  x\right)  \sim x^{2}$ for $x\ll1,$ the Taylor expansion of
$H^{a+\frac{1}{2}}\left(  x\right)  H^{\prime}\left(  x\right)  $ is given by
\begin{equation}
H^{a+\frac{1}{2}}\left(  x\right)  H^{\prime}\left(  x\right)  =\sum
_{j=0}^{\infty}d_{j}^{a}x^{j+2a+2},
\end{equation}
where $d_{0}^{a}=2$. Then one gets%
\begin{align}
\tilde{c}_{k}^{a}  &  =\int_{x_{1}}^{\Lambda}H^{a+\frac{1}{2}}\left(
x\right)  H^{\prime}\left(  x\right)  x^{-2k-1}dx+\int_{\delta}^{x_{1}}%
\sum_{j=0}^{\infty}d_{j}^{a}x^{j+2a+2}x^{-2k-1}dx\nonumber\\
&  =c_{k}^{a}-\sum_{j=0,j\neq2k-2a-2}^{\infty}\frac{d_{j}^{a}\delta
^{j+2\left(  a-k\right)  +2}}{j+2\left(  a-k\right)  +2}-\theta\left(
k-a-1\right)  d_{2k-2a-2}^{a}\ln\delta, \label{eq:Cak}%
\end{align}
where $\theta\left(  x\right)  $ is the Heaviside step function,
$0<x_{1}<\Lambda$, and $c_{k}^{a}$ is a constant independent of $L$.
Neglecting terms depending on $L$, one finds%
\begin{equation}
\tilde{c}_{k}^{a}\sim c_{k}^{a}-\theta\left(  k-a-1\right)  d_{2k-2a-2}^{a}%
\ln\frac{m_{p}\kappa u}{2\pi}. \label{eq:CakWithoutL}%
\end{equation}
Plugging eqn. $\left(  \ref{eq:CakWithoutL}\right)  $ into eqn. $\left(
\ref{eq:EntropyHx}\right)  $ gives us that the entropy near the horizon can be
written of form
\begin{equation}
S=\frac{C\left(  r_{h}^{2}\right)  }{4\pi^{2}m_{p}^{2}}%
{\displaystyle\sum\limits_{k=0}^{\infty}}
\left(  s_{k}+l_{k}\ln\frac{m_{p}\kappa}{2\pi}\right)  \left(  \frac
{m_{p}\kappa}{2\pi}\right)  ^{2k}.
\end{equation}
For $k=0$, we can choose $x_{1}=0$ in eqn. $\left(  \ref{eq:CakWithoutL}%
\right)  $ since $\int_{0}^{\Lambda}H^{a+\frac{1}{2}}\left(  x\right)
H^{\prime}\left(  x\right)  x^{-1}dx$ is convergent as $x\rightarrow0$. Hence,
one has%
\begin{equation}
c_{0}^{a}=\int_{0}^{\infty}H^{a+\frac{1}{2}}\left(  x\right)  H^{\prime
}\left(  x\right)  x^{-1}dx,
\end{equation}
and
\begin{equation}
s_{0}=\sum_{a=0}^{\infty}\frac{c_{0}^{a}\left(  \eta_{1}^{2}\right)  ^{a}%
\sqrt{\pi}}{\Gamma\left(  a+\frac{3}{2}\right)  }\int_{0}^{\infty}s^{\left(
l\right)  }\left(  u\right)  u^{l}du,
\end{equation}
where $\eta_{1}^{2}=\frac{3C_{3}^{2}}{16}-\frac{C_{4}}{4}$, and we use the
fact that $\xi_{l,k<n-l}^{n}=0,\xi_{l,0}^{l}=\left(  \eta_{1}^{2}\zeta_{0}%
^{1}\right)  ^{l}$, and $\zeta_{0}^{1}=1$. Since there is no $\ln\delta$ in
eqn. $\left(  \ref{eq:CakWithoutL}\right)  $ for $k=0$, we have $l_{0}=0$. For
$k=1$, only $\tilde{c}_{1}^{0}$ contributes to $l_{1}$, and we find%
\begin{equation}
l_{1}=-4f_{1}^{0}\int_{0}^{\infty}s\left(  u\right)  u^{2}du,
\end{equation}
where we have%
\[
f_{1}^{0}=\frac{r_{h}C^{\prime}\left(  r_{h}^{2}\right)  }{\kappa C\left(
r_{h}^{2}\right)  }\left[  1-\frac{C\left(  r_{h}^{2}\right)  f^{\prime\prime
}\left(  r_{h}\right)  }{4\kappa r_{h}C^{\prime}\left(  r_{h}^{2}\right)
}\right]  .
\]
Thus, we obtains for the entropy near horizon%
\begin{equation}
S\sim\frac{C\left(  r_{h}^{2}\right)  }{4\pi^{2}m_{p}^{2}}s_{0}-\frac{\kappa
r_{h}C^{\prime}\left(  r_{h}^{2}\right)  }{4\pi^{4}}\left[  1-\frac{C\left(
r_{h}^{2}\right)  f^{\prime\prime}\left(  r_{h}\right)  }{4\kappa
r_{h}C^{\prime}\left(  r_{h}^{2}\right)  }\right]  \int_{0}^{\infty}s\left(
u\right)  u^{2}du\ln m_{p}\kappa+\text{Finite terms as }\kappa m_{p}%
\rightarrow0\text{.} \label{eq:entropy-UV}%
\end{equation}

\subsection{Perturbative Case}

In this case, a wall near the horizon is needed to regulate the $x$-integrals.
As above, the function $H\left(  x\right)  $ can be presented in the form of
Taylor series%
\begin{equation}
H\left(  x\right)  =\sum_{i=2}^{\infty}C_{i}x^{i},
\end{equation}
where $C_{2}=1$. One then can have Taylor expansions for $H^{a+\frac{1}{2}%
}\left(  x\right)  H^{\prime}\left(  x\right)  $
\begin{equation}
H^{a+\frac{1}{2}}\left(  x\right)  H^{\prime}\left(  x\right)  =\sum
_{j=0}^{\infty}d_{j}^{a}x^{j+2a+2}, \label{eq:HaHprimeExpansion}%
\end{equation}
where $d_{0}^{a}=2$. The radial position of the brick wall near the horizon is
$r=r_{h}+r_{\varepsilon}$ $\left(  x=x_{\varepsilon}\right)  $. The invariant
distance of the wall from the horizon $\varepsilon$ is defined by%
\begin{equation}
\varepsilon=\int_{r_{h}}^{r_{h}+r_{\varepsilon}}\frac{dr}{\sqrt{f\left(
r\right)  }}=\int_{0}^{y_{\varepsilon}}\frac{2dy}{f^{\prime}\left(
f^{-1}\left(  y^{2}\right)  \right)  },
\end{equation}
where we define $y=\frac{T_{0}u}{xm_{p}}=\frac{\sqrt{f\left(  r\right)  }}{u}$
and $y_{\varepsilon}=\frac{T_{0}u}{x_{\varepsilon}m_{p}}$. Noting
$\frac{f^{\prime}\left(  f^{-1}\left(  0\right)  \right)  }{2\kappa}=1$, one
obtains
\begin{equation}
\varepsilon\kappa=\int_{0}^{y_{\varepsilon}}\frac{dy}{\frac{f^{\prime}\left(
f^{-1}\left(  y^{2}\right)  \right)  }{2\kappa}}=\int_{0}^{y_{\varepsilon}%
}\left(  1+%
{\displaystyle\sum\limits_{n=1}^{\infty}}
\tilde{f}_{n}y^{2n}\right)  dy=y_{\varepsilon}\left(  1+%
{\displaystyle\sum\limits_{n=1}^{\infty}}
\frac{\tilde{f}_{n}y_{\varepsilon}^{2n}}{2n+1}\right)  , \label{eq:upslion}%
\end{equation}
where we expand $\frac{2\kappa}{f^{\prime}\left(  f^{-1}\left(  y^{2}\right)
\right)  }$ in the integral and $\tilde{f}_{n}$ are coefficients of the
series. Solving eqn. $\left(  \ref{eq:upslion}\right)  $ for $y_{\varepsilon}$
gives
\begin{equation}
y_{\varepsilon}=\varepsilon\kappa\left(  1+%
{\displaystyle\sum\limits_{n=1}^{\infty}}
\zeta_{n}\left(  \varepsilon\kappa\right)  ^{2n}\right)  ,
\end{equation}
where $\zeta_{n}$ are determined by $\tilde{f}_{n}$. Using $x_{\varepsilon
}=\frac{T_{0}u}{y_{\varepsilon}m_{p}}$, one can relate $x_{\varepsilon}$ to
$\varepsilon$ by%
\begin{align}
x_{\varepsilon}^{a}  &  =\frac{T_{0}^{a}u^{a}}{m_{p}^{a}}\frac{1}{\left(
\varepsilon\kappa\right)  ^{a}}%
{\displaystyle\sum\limits_{n=0}^{\infty}}
\chi_{l}^{n}\left(  \varepsilon\kappa\right)  ^{2n}\text{,}\nonumber\\
\text{ }\ln\frac{m_{p}x_{\varepsilon}}{T_{0}u}  &  =%
{\displaystyle\sum\limits_{n=1}^{\infty}}
\chi_{0}^{n}\left(  \varepsilon\kappa\right)  ^{2n}-\ln\varepsilon\kappa,
\end{align}
where $\chi_{a}^{0}=1$. Focusing only on the near horizon contributions, we
neglect terms involving $L$ and use eqn. $\left(  \ref{eq:HaHprimeExpansion}%
\right)  $ to obtain
\begin{equation}
\int_{\delta}^{x_{\varepsilon}}H^{a+\frac{1}{2}}\left(  x\right)  H^{\prime
}\left(  x\right)  x^{-2k-1}dx\sim\sum_{j=0,j\neq2k-2a-2}^{\infty}\frac
{d_{j}^{a}x_{\varepsilon}^{j+2\left(  a-k\right)  +2}}{j+2\left(  a-k\right)
+2}+\theta\left(  k-a-1\right)  d_{2k-2a-2}^{a}\ln\frac{m_{p}x_{\varepsilon}%
}{T_{0}u}, \label{eq:Integral}%
\end{equation}
where for the logarithmic term we have $\int_{\delta}^{x_{\varepsilon}}%
x^{-1}dx=\ln\frac{x_{\varepsilon}}{\delta}=\ln\frac{x_{\varepsilon}m_{p}%
}{T_{0}u}-\frac{1}{2}\ln f\left(  L\right)  \sim\ln\frac{x_{\varepsilon}m_{p}%
}{T_{0}u}$. Plugging eqns. $\left(  \ref{eq:Integral}\right)  $ and $\left(
\ref{eq:Theta}\right)  $ into eqn. $\left(  \ref{eq:EntropyHx}\right)  $, one
finds for the entropy near the horizon that%
\begin{align}
S  &  \sim\frac{C\left(  r_{h}^{2}\right)  }{16\pi^{4}\varepsilon^{2}}%
\sum_{a=0}^{\infty}\sum_{j=0}^{\infty}\frac{\sqrt{\pi}a!}{\Gamma\left(
a+\frac{3}{2}\right)  \left(  2\pi\right)  ^{j+2a}}\left(  \frac{m_{p}%
}{\varepsilon}\right)  ^{j+2a}\nonumber\\
&  \text{ }%
{\displaystyle\sum\limits_{k=0,k\neq\frac{j}{2}+a+1}^{\infty}}
{\displaystyle\sum\limits_{n=0}^{\infty}}
{\displaystyle\sum\limits_{p=0}^{\infty}}
\frac{d_{j}^{a}f_{k}^{a}\chi_{j+2\left(  a-k\right)  +2}^{n}}{j+2\left(
a-k\right)  +2}\left(  \frac{m_{p}\kappa}{2\pi}\right)  ^{2p}\left(
\kappa^{2}\varepsilon^{2}\right)  ^{k+n}%
{\displaystyle\sum\limits_{q=0}^{p+a}}
\frac{\xi_{a,p}^{q}}{q!}\int_{0}^{\infty}u^{j+2l+2p+q+2}s^{\left(  q\right)
}\left(  u\right)  du\nonumber\\
&  +\frac{C\left(  r_{h}^{2}\right)  \kappa^{2}}{16\pi^{4}}%
{\displaystyle\sum\limits_{n=0}^{\infty}}
\left(  \kappa^{2}\varepsilon^{2}\right)  ^{n}\sum_{a=0}^{\infty}\sum
_{j=0}^{\infty}%
{\displaystyle\sum\limits_{p=0}^{\infty}}
\frac{\sqrt{\pi}a!}{\Gamma\left(  a+\frac{3}{2}\right)  }d_{2j}^{a}%
f_{j+a+1}^{a}\chi_{0}^{n}\left(  \frac{m_{p}\kappa}{2\pi}\right)
^{2j+2a+2p}\nonumber\\
&  \text{ \ \ \ \ \ }%
{\displaystyle\sum\limits_{q=0}^{p+a}}
\xi_{a,p}^{q}\int_{0}^{\infty}u^{2j+2a+2p+q+2}\frac{s^{\left(  q\right)
}\left(  u\right)  }{q!}du\nonumber\\
&  -\ln\left(  \kappa\varepsilon\right)  \frac{C\left(  r_{h}^{2}\right)
\kappa^{2}}{16\pi^{4}}\sum_{a=0}^{\infty}\sum_{j=0}^{\infty}%
{\displaystyle\sum\limits_{p=0}^{\infty}}
\frac{\sqrt{\pi}a!}{\Gamma\left(  a+\frac{3}{2}\right)  }d_{2j}^{a}%
f_{j+a+1}^{a}\left(  \frac{m_{p}\kappa}{2\pi}\right)  ^{2j+2a+2p}\nonumber\\
&  \text{ \ \ \ \ }%
{\displaystyle\sum\limits_{q=0}^{p+a}}
\frac{\xi_{a,p}^{q}}{q!}\int_{0}^{\infty}s^{\left(  q\right)  }\left(
u\right)  u^{2j+2a+2p+q+2}du. \label{eq:EntropyNearP}%
\end{align}
At first sight, it seems impossible to single out the most divergent part of
eqn. $\left(  \ref{eq:EntropyNearP}\right)  $ since $j+2a$ in the first term
of eqn. $\left(  \ref{eq:EntropyNearP}\right)  $ can go to infinity. However,
the brick wall is put at $r=r_{h}+r_{\varepsilon}$ to cut off some unknown
quantum physics of gravity. In this sense, the invariant distance of the wall
from the horizon $\varepsilon$ could be given by $\varepsilon\sim m_{p}$.
Indeed in the 't Hooft's original calculation for Schwarzschild black holes,
requiring that the entropy of the radiation near the horizon $S_{Brick}$
$\lesssim$ the Black hole's Bekenstein-Hawking entropy $S_{BH}$ also gives
$\varepsilon\gtrsim\sqrt{\frac{1}{90\pi}}m_{p}$ for a scalar field. Thus, we
define $\alpha$ such as $\varepsilon=\alpha m_{p}$. Replacing $\varepsilon$ by
$\alpha m_{p}$ in eqn. $\left(  \ref{eq:EntropyNearP}\right)  $, we find for
the entropy%
\begin{equation}
S\sim\frac{C\left(  r_{h}^{2}\right)  }{4\pi^{2}m_{p}^{2}}s_{0}-\frac{\kappa
r_{h}C^{\prime}\left(  r_{h}^{2}\right)  }{4\pi^{4}}\left[  1-\frac{C\left(
r_{h}^{2}\right)  f^{\prime\prime}\left(  r_{h}\right)  }{4\kappa
r_{h}C^{\prime}\left(  r_{h}^{2}\right)  }\right]  \int_{0}^{\infty}s\left(
u\right)  u^{2}du\ln\kappa m_{p}+\text{Finite terms as }m_{p}\kappa
\rightarrow0, \label{eq:entropy-pt}%
\end{equation}
where we define%
\begin{equation}
s_{0}=\frac{1}{4\pi^{2}\alpha^{2}}\sum_{a=0}^{\infty}\sum_{j=0}^{\infty}%
\frac{\sqrt{\pi}a!}{\Gamma\left(  a+\frac{3}{2}\right)  }\left(  \frac{1}%
{2\pi\alpha}\right)  ^{j+2a}\frac{d_{j}^{a}}{j+2a+2}\left(
{\displaystyle\sum\limits_{q=0}^{a}}
\frac{\xi_{a,0}^{q}}{q!}\int_{0}^{\infty}u^{j+2a+q+2}s^{\left(  q\right)
}\left(  u\right)  du\right)  . \label{eq:Rs0}%
\end{equation}

\subsection{Discussion}

For a massless scalar field, we find the entropy near horizon in both cases
can be written as%
\begin{equation}
S\sim\frac{As_{0}}{16\pi^{3}m_{p}^{2}}+s_{L}\ln m_{p}\kappa+\text{Finite terms
as }\kappa m_{p}\rightarrow0\text{,} \label{eq:Entropy-MDR}%
\end{equation}
where $s_{L}=-\frac{\kappa r_{h}C^{\prime}\left(  r_{h}^{2}\right)  }%
{45}\left[  1-\frac{C\left(  r_{h}^{2}\right)  f^{\prime\prime}\left(
r_{h}\right)  }{4\kappa r_{h}C^{\prime}\left(  r_{h}^{2}\right)  }\right]  $,
and $A=4\pi C\left(  r_{h}^{2}\right)  $ is the horizon area. For the scenario
without the MDR, the entropy near horizon
\cite{EBW-'tHooft:1984re,EBW-Solodukhin:1994yz,EBW-Solodukhin:2011gn} is
\begin{equation}
S\sim\frac{A}{360\alpha^{2}\pi m_{p}^{2}}+s_{L}\ln m_{p}\kappa+\text{Finite
terms as }\kappa m_{p}\rightarrow0, \label{eq:Entropy-No-MDR}%
\end{equation}
where we let the proper distance $\varepsilon=\alpha m_{p}$. On the other
hand, the first law of black hole thermodynamics $dS_{B}=\frac{dM}{T}$, and
eqn. $\left(  \ref{eq:Temp-HJ}\right)  $ leads to the modified entropy of the
black hole%
\begin{equation}
S_{B}\sim\frac{A}{4m_{p}^{2}}+\frac{\pi}{8}\left(  7C_{3}^{2}-20C_{4}\right)
\ln\kappa m_{p}+\text{Finite terms as }m_{p}\kappa\rightarrow0,
\label{eq:entropy-BH}%
\end{equation}
where $A=16\pi M^{2}$ and $\kappa=\frac{1}{4M}.$ As in the brick wall model
originally introduced by 't Hooft, we could adjust $\alpha$ ($s_{0}$ via eqn.
$\left(  \ref{eq:Rs0}\right)  $) to make the leading term in eqn. $\left(
\ref{eq:Entropy-MDR}\right)  $ the same as that in eqn. $\left(
\ref{eq:entropy-BH}\right)  $. Moreover, the subleading logarithmic term in
eqn. $\left(  \ref{eq:Entropy-MDR}\right)  $ only depends on the properties of
the black hole. However for $S_{B}$, the subleading logarithmic term depends
both on the black hole and MDR ($C_{3}$ and $C_{4}$). This observation might
suggest that the explanations for statistical origin of the black holes'
entropy need more than the entropy of a thermal gas of Hawking particles near
the horizon.

Since the deformed Hamilton-Jacobi equations and the corrections to the
Hawking temperature are same for fermions and scalars with the same MDR, one
may wonder if eqns. $\left(  \ref{eq:entropy-UV}\right)  $ and $\left(
\ref{eq:entropy-pt}\right)  $ also work for fermions. In fact, it has been
shown in \cite{EBW-Iizuka:2013kma} that the same argument in this section held
for fermions if an appropriate boundary condition was taken instead of the too
restrictive Dirichlet boundary condition.

For a MDR with $H\left(  x\right)  $ in the UV finite case, we have shown that
a brick wall near the horizon is not needed since the entropy is finite as one
approaches the horizon. However, if one expands $H\left(  x\right)  $ as a
power series of $x$ and calculates the entropy in the perturbative case, it
seems that a wall near the horizon is needed to regulate the divergence. How
can we reconcile the contradiction? As noted in \cite{EBW-Wang:2015bwa}, the
divergence of the entropy in the perturbative case as $\alpha\rightarrow0$ is
more like due to the breaking down of the Taylor series. For the typical
energy $\omega\sim T_{0}=\frac{\hbar\kappa}{2\pi}\,$, one finds that $H\left(
x\right)  $ and the MDR corrections to the entropy are powers of $\frac
{\omega}{\sqrt{f\left(  r\right)  }m_{p}}\sim\frac{\hbar\kappa}{2\pi
\sqrt{f\left(  r\right)  }m_{p}}$. At the wall at $r_{\varepsilon}\approx
r_{h}+2\kappa m_{p}^{2},$ we have $\frac{\hbar\kappa}{\sqrt{f\left(
r_{\varepsilon}\right)  }m_{p}}\sim\frac{1}{4\pi}$. Thus, the perturbative
case is valid outside the wall at $r_{\varepsilon}=r_{h}+2\kappa m_{p}^{2}$.
However, the perturbation would break down deep within the wall, and the
closed form of $H\left(  x\right)  $ is needed.

\section{Black Hole Evaporation}

\label{Sec:BHE}

In \cite{BHE-Page:1976df}, Page counted the number of modes per frequency
interval with periodic boundary conditions in a large container around the
black hole and divided it by the time it takes a particle to cross the
container. He then related the expected number emitted per mode $n$ to the
average emission rate per frequency interval $\frac{dn}{dt}$ by%
\begin{equation}
\frac{dn}{dt}=n\frac{d\omega}{2\pi\hbar}, \label{eq:DnDt}%
\end{equation}
for each mode and frequency interval $\left(  \omega,\omega+d\omega\right)  $.
Following the same argument, we find that in the MDR case
\begin{equation}
\frac{dn}{dt}=n\frac{\partial\omega}{\partial p_{r}}\frac{dp_{r}}{2\pi\hbar
}=n\frac{d\omega}{2\pi\hbar}, \label{eq:DnDt-MDR}%
\end{equation}
where $\frac{\partial\omega}{\partial p_{r}}$ is the radial velocity of the
particle, and the number of modes between the wavevector interval $\left(
p_{r},p_{r}+dp_{r}\right)  $ is $\frac{dp_{r}}{2\pi\hbar}$. Since each
particle carries off the energy $\omega$, the total luminosity is obtained
from multiplying $\frac{dn}{dt}$ by the energy $\omega$ and summing up over
all energy $\omega$ and quantum numbers, denoted by $i$,%
\begin{equation}
L=%
{\displaystyle\sum\limits_{i}}
\int\omega n_{\omega,i}\frac{d\omega}{2\pi\hbar}.
\end{equation}
However, some of the radiation emitted by the horizon might not be able to
reach the asymptotic region. Before the radiation reaches the distant
observer, they must pass the curved spacetime around the black hole horizon,
which plays the role of a potential barrier. This effect on $L$ can be
described by a greybody factor from the scattering coefficients of the black
hole. Actually, the greybody factor is given by $\left\vert T_{i}\left(
\omega\right)  \right\vert ^{2}$, where $T_{i}\left(  \omega\right)  $
represents the transmission coefficient of the black hole barrier which in
general can depend on the energy $\omega$ and quantum numbers $i$ of the
particle. Taking the greybody factor into account, we find for the total
luminosity that%
\begin{equation}
L=%
{\displaystyle\sum\limits_{i}}
\int\left\vert T_{i}\left(  \omega\right)  \right\vert ^{2}\omega n_{\omega
,i}\frac{d\omega}{2\pi\hbar}\text{\thinspace}. \label{eq:Luminosity}%
\end{equation}
Since the relevant radiation usually have the energy of order $\hbar M^{-1}$,
where $M$ is the mass of the black hole, one should use the wave equations
given in the appendix to compute $\left\vert T_{i}\left(  \omega\right)
\right\vert ^{2}$ accurately. However, solving the wave equations for
$\left\vert T_{i}\left(  \omega\right)  \right\vert ^{2}$ could be very
complicated. On the other hand, one can use the geometric optics approximation
to estimate $\left\vert T_{i}\left(  \omega\right)  \right\vert ^{2}$. In the
geometric optics approximation, we assume $\omega\gg M$, and high energy waves
will be absorbed unless they are aimed away from the black hole. Hence we have
$\left\vert T_{i}\left(  \omega\right)  \right\vert ^{2}=1$ for all the
classically allowed energy $\omega$ and quantum numbers $i$ of the particle,
while $\left\vert T_{i}\left(  E\right)  \right\vert ^{2}=0$ otherwise. For
the usual dispersion relations, the Stefan's law for black holes is obtained
in this approximation. In the remaining of the section, we will discuss
evaporations of a 4D spherically symmetric black hole with the mass $M\gg
m_{p}$ and a 2D black hole. For simplicity, we assume that the particles are
massless and neutral.

\subsection{4D Spherically Symmetric Black Hole}

To find the classically allowed values of angular momentum $l$ with fixed
value of energy $\omega$, we consider eqn. $\left(  \ref{eq:Pr}\right)  $ for
a massless particle in a 4D spherically symmetric black hole, where we have
$\lambda=\left(  l+\frac{1}{2}\right)  ^{2}\hbar^{2}$. Since $p_{r}$ is always
a real number in the geometric optics approximation, one has an upper bound on
$\lambda$%
\begin{equation}
\lambda\leq C\left(  r^{2}\right)  m_{p}^{2}H\left(  \frac{\omega}{m_{p}%
\sqrt{f\left(  r\right)  }}\right)  .
\end{equation}
Suppose $C\left(  r^{2}\right)  m_{p}^{2}H\left(  \frac{\omega}{m_{p}%
\sqrt{f\left(  r\right)  }}\right)  $ has a minimum at $r_{\min}$ and this
minimum is denoted by $\lambda_{\max}$. If the particles overcome the angular
momentum barrier and get absorbed by the black hole, one must have
$\lambda\leq\lambda_{\max}$. Thus, the eqn. $\left(  \ref{eq:Luminosity}%
\right)  $ becomes%
\begin{align}
L  &  =g_{s}\int\frac{\omega d\omega}{2\pi\hbar^{3}}\int_{0}^{\lambda_{\max}%
}n\left(  \frac{\omega\left(  1+\Delta\right)  }{T_{0}}\right)  d\left[
\left(  l+\frac{1}{2}\right)  ^{2}\hbar^{2}\right] \nonumber\\
&  =\frac{g_{s}C\left(  r_{h}^{2}\right)  m_{p}^{2}}{2\pi\hbar^{3}}\int\omega
d\omega\int_{0}^{\frac{C\left(  r_{\min}^{2}\right)  }{C\left(  r_{h}%
^{2}\right)  }H\left(  \frac{\omega}{m_{p}\sqrt{f\left(  r_{\min}\right)  }%
}\right)  }n\left(  \frac{\omega\left(  1+\Delta\right)  }{T_{0}}\right)  dz,
\label{eq:L-Delta}%
\end{align}
where $g_{s}$ is the number of polarization, $z=\frac{\left(  l+\frac{1}%
{2}\right)  ^{2}\hbar^{2}}{C\left(  r_{h}^{2}\right)  m_{p}^{2}}$, and we use
eqn. $\left(  \ref{eq:Entropy}\right)  $ for $n_{\omega,l}$. Defining
$n_{a}\left(  u\right)  $ by
\begin{equation}
n_{a}\left(  u\right)  =%
{\displaystyle\sum\limits_{k=0}^{\infty}}
\left(  \frac{T_{0}u}{m_{p}}\right)  ^{2k}%
{\displaystyle\sum\limits_{n=0}^{k+a}}
\left[  \frac{n^{\left(  n\right)  }\left(  u\right)  u^{n}}{n!}\right]
\xi_{a,k}^{n}\text{,} \label{eq:na}%
\end{equation}
we find
\begin{equation}
n\left(  \frac{\omega\left(  1+\Delta\right)  }{T_{0}}\right)  =\sum
_{a=0}^{\infty}z^{a}n_{a}\left(  u\right)  , \label{eq:N-z}%
\end{equation}
where $\xi_{a,k}^{n}$ is given by eqn. $\left(  \ref{eq:Xsi}\right)  $.
Substituting eqn. $\left(  \ref{eq:N-z}\right)  $ into eqn. $\left(
\ref{eq:L-Delta}\right)  $ and integrating eqn. $\left(  \ref{eq:L-Delta}%
\right)  $ over $z$ gives
\begin{equation}
L=\frac{g_{s}C\left(  r_{h}^{2}\right)  T_{0}^{2}}{2\pi\hbar^{3}}\sum
_{a=0}^{\infty}\frac{m_{p}^{2}}{\left(  a+1\right)  }\frac{C^{a+1}\left(
r_{\min}^{2}\right)  }{C^{a+1}\left(  r_{h}^{2}\right)  }\int_{0}^{\infty
}un_{a}\left(  u\right)  H^{a+1}\left(  \frac{T_{0}u}{m_{p}\sqrt{f\left(
r_{\min}\right)  }}\right)  du, \label{eq:L-H}%
\end{equation}
where we let $u_{\max}=\infty$ for $M\gg m_{p}$. Since $H\left(  x\right)
=x^{2}+\sum\limits_{n\geq3}C_{n}x^{n}$, we define $h_{m}^{a}$ by
\begin{equation}
H^{a}\left(  x\right)  =x^{2a}%
{\displaystyle\sum\limits_{m=0}^{\infty}}
h_{m}^{a}x^{m}, \label{eq:Ha}%
\end{equation}
where $h_{0}^{a}=1,h_{1}^{a}=aC_{3}$ and $h_{2}^{a}=C_{4}a+\frac{C_{3}%
^{2}\left(  a-1\right)  a}{2}$. Plugging eqns. $\left(  \ref{eq:Ha}\right)  $
and $\left(  \ref{eq:na}\right)  $ into eqn. $\left(  \ref{eq:L-H}\right)  $
gives
\begin{equation}
L=\frac{g_{s}C\left(  r_{h}^{2}\right)  T_{0}^{4}}{2\pi\hbar^{3}}%
{\displaystyle\sum\limits_{j=0}^{\infty}}
\left(  \frac{T_{0}}{m_{p}}\right)  ^{j}\sum_{a=0}^{\left[  \frac{j}%
{2}\right]  }\frac{1}{a+1}\left[
{\displaystyle\sum\limits_{k=0}^{\left[  \frac{j}{2}\right]  -a}}
h_{j-2a-2k}^{a+1}\left(
{\displaystyle\sum\limits_{i=0}^{k+a}}
\frac{\xi_{a,k}^{i}}{i!}N_{i,j,l,k}\right)  \right]  , \label{eq:L-N}%
\end{equation}
where $\left[  x\right]  =\max\left\{  m\in Z\text{ }|\text{ }m\leq x\right\}
$ and we define%
\begin{equation}
N_{i,j,a,k}=\int_{0}^{\infty}\frac{n^{\left(  i\right)  }\left(  u\right)
u^{j+i+3}}{f^{\frac{j}{2}-k+1}\left(  r_{\min}\right)  }\frac{C^{a+1}\left(
r_{\min}^{2}\right)  }{C^{a+1}\left(  r_{h}^{2}\right)  }du. \label{eq:Nnpak}%
\end{equation}

We now use eqn. $\left(  \ref{eq:L-N}\right)  $ to calculate the luminosity in
the Schwarzschild metric to $\mathcal{O}\left(  \frac{m_{p}^{2}}{M^{2}%
}\right)  $. For the Schwarzschild metric, one has $f\left(  r\right)
=1-\frac{2M}{r},$ $r_{h}=2M$, $C\left(  r^{2}\right)  =r^{2}$ and
$\kappa=\frac{1}{4M}$. Taking the derivative of $C\left(  r^{2}\right)
m_{p}^{2}H\left(  \frac{\omega}{m_{p}\sqrt{f\left(  r\right)  }}\right)  $ and
equating it to zero, one finds
\begin{align}
r_{\min}  &  =3M\left(  1+\frac{\sqrt{3}C_{3}}{6}\frac{T_{0}u}{m_{p}}%
+\frac{12C_{4}-7C_{3}^{2}}{12}\frac{T_{0}^{2}u^{2}}{m_{p}^{2}}+\mathcal{O}%
\left(  \frac{T_{0}^{3}}{m_{p}^{3}}\right)  \right)  ,\nonumber\\
\lambda_{\max}  &  =27M^{2}u^{2}T_{0}^{2}\left(  1+\sqrt{3}C_{3}\frac{T_{0}%
u}{m_{p}}+\frac{12C_{4}-C_{3}^{2}}{4}\frac{T_{0}^{2}u^{2}}{m_{p}^{2}%
}+\mathcal{O}\left(  \frac{T_{0}^{3}}{m_{p}^{3}}\right)  \right)  .
\label{eq:R amd lamda min}%
\end{align}
For emitting $n_{s}$ species of massless scalars and $n_{f}$ species of
massless spin-$1/2$ fermions from a Schwarzschild black hole into empty space,
putting eqns. $\left(  \ref{eq:R amd lamda min}\right)  $ into eqn. $\left(
\ref{eq:L-N}\right)  $ gives the total luminosity
\begin{align}
L  &  =\frac{9m_{p}^{2}}{40960\pi M^{2}}\left\{  \left(  n_{s}+\frac{7}%
{4}n_{f}\right)  +\left(  0.26n_{s}+0.50n_{f}\right)  C_{3}\frac{m_{p}}%
{M}\right. \nonumber\\
&  \left.  +\left[  \left(  0.15n_{s}+0.29n_{f}\right)  C_{3}^{2}-\left(
0.24n_{s}+0.46n_{f}\right)  C_{4}\right]  \frac{m_{p}^{2}}{M^{2}}%
+\mathcal{O}\left(  \frac{m_{p}^{3}}{M^{3}}\right)  \right\}  \text{.}
\label{eq:L-SC}%
\end{align}
The perturbative parameter $\omega/m_{p}$ appears in calculating the total
luminosity. For massless particles, we could have estimations%
\begin{equation}
\omega\sim T_{0}\sim\frac{m_{p}^{2}}{8\pi M}.
\end{equation}
Thus, this perturbative parameter becomes%
\begin{equation}
\frac{\omega}{m_{p}}\sim\frac{m_{p}}{M},
\end{equation}
which explains why the the total luminosity in eqn. $\left(  \ref{eq:L-SC}%
\right)  $ is suppressed by powers of $m_{p}/M$.

In the geometric optics approximation, the Schwarzschild black hole can be
described as a black sphere for absorbing particles. The total luminosity are
determined by the radius of the black sphere $R$ and the temperature of the
black hole $T$. Note that one has $R=\sqrt{\frac{\lambda_{\max}}{\omega^{2}}}$
and $T_{eff}\approx T_{0}\left(  1-\Delta\right)  $, where for massless
particles
\begin{equation}
\Delta=\frac{1}{32m_{p}^{2}}\left[  \left(  4C_{4}-3C_{3}^{2}\right)
\frac{\lambda}{2M^{2}}+8\left(  4C_{4}-C_{3}^{2}\right)  \omega^{2}\right]  .
\end{equation}
Consider a sub-luminal case with $C_{3}=0$ and $C_{4}>0$, where the total
luminosity decreases due to the MDR effects. In this case, the MDR effects
increase the radius of the black sphere while they decrease the temperature of
the black hole. The competition between the increased radius and the decreased
temperature determines whether the luminosity would increase or decrease. It
appears form eqn. $\left(  \ref{eq:L-SC}\right)  $ that the effects of
decreased temperature wins the competition.

\subsection{2D Black Hole}

Suppose the metric of a 2D black hole is given by%
\begin{equation}
ds^{2}=f\left(  r\right)  dt^{2}-\frac{1}{f\left(  r\right)  }dr^{2},
\end{equation}
where $f\left(  r\right)  $ has a simple zero at $r=r_{h}$. Here we consider a
neutral and massless scalar particle governed by the modified dispersion
relation
\begin{equation}
E^{2}=p^{2}-\frac{Cp^{4}}{m_{p}^{2}},
\end{equation}
which is the Corley and Jacobson dispersion relation for $C>0$
\cite{IN-Corley:1996ar}. Expressing $p$ in terms of $E$ gives%
\begin{equation}
p^{2}=E^{2}-\frac{CE^{4}}{m_{p}^{2}}+\mathcal{O}\left(  \frac{E^{4}}{m_{p}%
^{4}}\right)  .
\end{equation}
For the 2D black hole with the event horizon at $r=r_{h}$, eqn. $\left(
\ref{eq:delta}\right)  $ gives
\begin{equation}
\Delta=-\frac{f^{\prime\prime}\left(  r_{h}\right)  }{8\kappa^{2}}\frac
{\omega^{2}}{m_{p}^{2}}C+\mathcal{O}\left(  \frac{\omega^{4}}{m_{p}^{4}%
}\right)  , \label{eq:2DDelta}%
\end{equation}
where we use $m=0$, $\tilde{\omega}\left(  r_{h}\right)  =\omega$, $\lambda
=0$, $\alpha=1$, $\gamma=0$, $C_{3}=0$, and $C_{4}=C$. For $\omega
<\omega_{\max}$, the term $\frac{\omega^{2}C}{m_{p}^{2}}$ in eqn. $\left(
\ref{eq:2DDelta}\right)  $ dominates, and hence the terms $\mathcal{O}\left(
\frac{\omega^{4}}{m_{p}^{4}}\right)  $ are neglected for simplicity. Define
$\eta=-\frac{f^{\prime\prime}\left(  r_{h}\right)  }{8\kappa^{2}}$ which
becomes $1$ for a 2D Schwarzschild black hole with $f\left(  r\right)
=1-\frac{2M}{r}$. In this case, we can choose the cutoff of the effective
theories $\Lambda=\frac{\alpha m_{p}}{\sqrt{\left\vert \eta C\right\vert }}$,
where $0<\alpha<1$. Note that $\left\vert \Delta\right\vert <1$ for
$\omega<\Lambda.$ Therefore, the luminosity for the black hole is%
\begin{align}
L  &  =\int_{0}^{\omega_{\max}}\omega n\left[  \frac{\omega}{T_{0}}\left(
1+\frac{\omega^{2}\eta C}{m_{p}^{2}}\right)  \right]  \frac{d\omega}{2\pi
\hbar}\nonumber\\
&  =\frac{\kappa^{2}m_{p}^{2}}{8\pi^{3}}\int_{0}^{u_{\max}}un\left[  u\left(
1+\frac{\eta C\kappa^{2}m_{p}^{2}u^{2}}{4\pi^{2}}\right)  \right]  du,
\label{eq:L-2D}%
\end{align}
where $T_{0}=\frac{\hbar\kappa}{2\pi}$ and $u_{\max}=\min\left\{  \frac{2\pi
M}{\kappa m_{p}^{2}},\frac{\alpha}{\sqrt{\left\vert \eta C\right\vert }}%
\frac{2\pi}{m_{p}\kappa}\right\}  $. For $\kappa m_{p}\ll1$, we can let
$u_{\max}=\infty$ and then find%
\begin{equation}
L=\frac{\kappa^{2}m_{p}^{2}}{48\pi}\left(  1-\frac{2\eta C}{5}\kappa^{2}%
m_{p}^{2}+\mathcal{O}\left(  \kappa^{4}m_{p}^{4}\right)  \right)  .
\label{eq:2DLuminosityLargeM}%
\end{equation}
For $M<\frac{\alpha m_{p}}{\sqrt{\left\vert \eta C\right\vert }},$ we have
$u_{\max}=\frac{2\pi M}{\kappa m_{p}^{2}}$ and find%
\begin{equation}
L=\frac{\kappa M}{4\pi^{2}}\left[  1-\frac{\pi\kappa M}{2\kappa^{2}m_{p}^{2}%
}\left(  1+\frac{2\eta C\kappa M}{3\pi}\right)  +\mathcal{O}\left(
\kappa^{-4}m_{p}^{-4}\right)  \right]  . \label{eq:2DLuminositySmallM}%
\end{equation}
From eqn. $\left(  \ref{eq:2DLuminosityLargeM}\right)  $ for small $\kappa
m_{p}$ and eqn. $\left(  \ref{eq:2DLuminositySmallM}\right)  $ for large
$\kappa m_{p}$, we can conclude that the coefficients $C$ impacts the black
hole's luminosity only in the intermediate range of $\kappa m_{p}$ noticeably.

\begin{figure}[tb]
\begin{centering}
\includegraphics[scale=1.5]{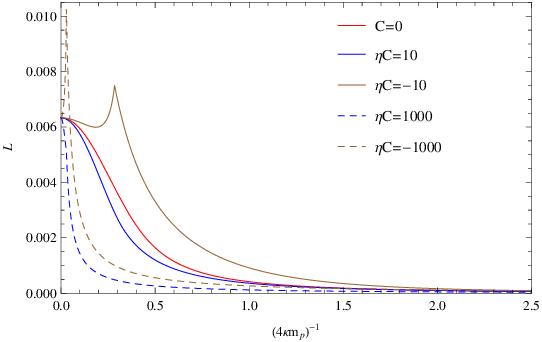}
\par\end{centering}
\caption{The luminosity $L$ of a 2D black hole against $(4\kappa m_{p})^{-1}$
with $\alpha=0.9$. }%
\label{fig:LC}%
\end{figure}

For the intermediate range, FIG. \ref{fig:LC} plots the luminosity $L$ against
$\left(  4\kappa m_{p}\right)  ^{-1}$, which becomes $\frac{m_{p}}{M}$ for a
2D Schwarzschild black hole with the mass $M$. In FIG. \ref{fig:LC}, we have
$\alpha=0.9$ and $M=\frac{1}{4\kappa}$ for $u_{\max}$. We plot $L$ vs $\left(
4\kappa m_{p}\right)  ^{-1}$ in FIG. \ref{fig:LC} for the usual case with
$C=0$ (red line), the ones with $\eta C=10$ and $1000$ (solid and dashed blue
lines, respectively), and the ones with $\eta C=-10$ and $-1000$ (solid and
dashed brown lines, respectively). For the $\eta C<0$ cases, there are "weird"
peaks in FIG. \ref{fig:LC}, which are due to the transition from $\frac{2\pi
M}{\kappa m_{p}^{2}}$ to $\frac{\alpha}{\sqrt{\left\vert \eta C\right\vert }%
}\frac{2\pi}{m_{p}\kappa}$ in $u_{\max}$. However, such transitions is barely
seen for the $\eta C>0$ cases. In our calculations, the luminosities are
determined not only by the modified Hawking temperature but also the range of
integration of $u$ in eqn. $\left(  \ref{eq:L-2D}\right)  $. When
$M>\frac{\alpha m_{p}}{\sqrt{\left\vert \eta C\right\vert }}$, in the $\eta
C<0$/$\eta C>0$ cases the ranges of integration are less than that in the
usual case, which tends to decrease the luminosity. In the $\eta C>0$ cases,
it shows from eqn. $\left(  \ref{eq:2DDelta}\right)  $ that the modified
Hawking temperatures are lower than that in the usual case. Thus, the
luminosities $L$ become smaller due to the decreased temperature and the
shrunken range. From eqn. $\left(  \ref{eq:2DDelta}\right)  ,$ we find that
the modified Hawking temperatures in the $\eta C<0$ cases are higher than that
in the usual case. Thus, the competition between the increased temperature and
the shrunken range determines the luminosity. The effect of the increased
temperature dominates over that of the shrunken range for $\eta C=-10$ and
vice versa for $\eta C=-1000$.

\begin{figure}[tb]
\begin{centering}
\includegraphics[scale=1.5]{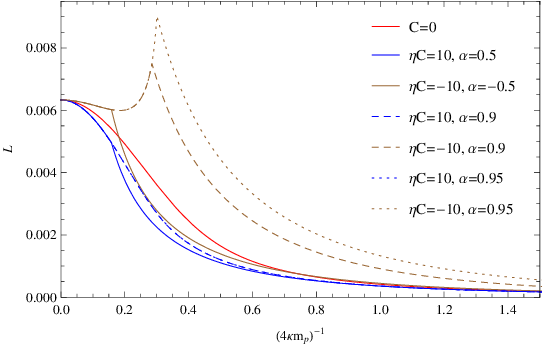}
\par\end{centering}
\caption{The luminosity $L$ of a 2D black hole against $\left(  4\kappa
m_{p}\right)  ^{-1}$ with $a=0.5,$ $0.9$ and $0.95$.}%
\label{fig:La}%
\end{figure}

To see how the luminosities $L$ depend on values of $\alpha$, we plot $L$ vs
$\left(  4\kappa m_{p}\right)  ^{-1}$ in FIG. \ref{fig:La} for the usual case
with $C=0$ (red line), the ones with $\eta C=10$ (blue lines), and the ones
with $\eta C=-10$ (brown lines) with $\alpha=0.5$ (solid lines)$,$ $0.9$
(dashed lines), and $0.95$ (dotted lines). Note that $\alpha$ parameterizes
the unknown quantum gravity ultraviolet cutoff $\Lambda$. In FIG.
\ref{fig:La}, it suggests that the $\eta C<0$ cases are highly sensitive to
the physics at high energies while the $\eta C>0$ ones are not. If $\eta>0$,
$\eta C<0$/$\eta C>0$ implies that the particles are super-/sub-luminal. The
author in \cite{BHE-Jacobson:2001kz} has shown that the Hawking radiation with
sub-luminal dispersion was not sensitive to Lorentz violation at high energies
due to the "mode conversion". However, the outgoing black hole modes with
super-luminal dispersion emanated from some unknown quantum gravity processes.

\section{Discussion and Conclusion}

\label{Sec:Con}

In this paper, we used the Hamilton-Jacobi method to investigate the effects
of the MDR on the Hawking radiation. Our results suggest that the thermal
spectrum of radiations near horizon is robust. In fact, if the difference
between the modified dispersion relation and the relativistic one was
suppressed by the fundamental energy scale $m_{p},$ we found that the
deviation of the effective Hawking temperature from the standard one was also
suppressed by $m_{p}$. For a particle with the typical energy $\omega\sim
\frac{m_{p}^{2}}{M}$, the deviation was given by powers of $\frac{m_{p}^{2}%
}{M^{2}}$.\ Nevertheless, there are some potential corrections to the
effective Hawking temperature which are not included in our calculations:

\begin{enumerate}
\item[$\left(  a\right)  $] Back-reaction effects which occurs at order
$\frac{\omega}{M}$. For a particle with $\omega\sim\frac{m_{p}^{2}}{M}$, they
are order of $\frac{m_{p}^{2}}{M^{2}}$. However, the Hamilton-Jacobi method is
incapable of computing them since the metric is fixed in this method. On the
other hand, back-reaction appears in the null geodesic method
\cite{IN-Kraus:1994by,IN-Kraus:1994fj} to ensure energy conservation during
the emission of a particle via tunneling through the horizon. These
corrections lead to non-thermal corrections to the black-hole radiation
spectrum. Note that there are some attempts to incorporate back-reaction
effects into the Hamilton-Jacobi method using the rainbow metric
\cite{CON-Medved:2005yf,CON-Ding:2013bz}.

\item[$\left(  b\right)  $] Higher order WKB corrections. In the
Hamilton-Jacobi method, we take the semi-classical limit $\hbar\rightarrow0$
and keep only leading order terms to calculate the Hawking temperature.
Therefore, one may wonder if the Hawking temperature could receive higher
order corrections in $\hbar$ beyond the semiclassical one. The corrections has
been estimated in \cite{CON-Banerjee:2008cf} and was given by powers of
$\frac{m_{p}^{2}}{M^{2}}$. However for the usual case, several authors
\cite{CON-Yale:2010tn,CON-Chatterjee:2009mw,CON-Wang:2009zzw} argued that the
tunneling method yielded no higher-order corrections to the Hawking
temperature. Whether such arguments also work for the MDR cases needs to be checked.
\end{enumerate}

In this paper, we first used the Hamilton-Jacobi method to calculate tunneling
rates of radiations across the horizon and the effective Hawking
temperatures.\ After the spectrum of radiations near the horizon was obtained,
the thermal entropy of radiations near the horizon and the luminosity of the
black hole were computed. Our main results are as follows:

\begin{itemize}
\item In section \ref{Sec:DHJE} and the appendix, we used heuristic arguments
and effective field theories, respectively to derive the deformed
Hamilton-Jacobi equations incorporate the MDR with the static preferred frame.
Note that these methods can easily be generalized to any preferred frame.

\item In section \ref{Sec:DHJE}, the deformed Hamilton-Jacobi equations was
solved for $\partial_{r}I$, and the imaginary part of $I$ was obtained by
computing the residue of $\partial_{r}I$ at $r=r_{h}$. The assumption for our
calculation was also discussed, which required that the singularity structure
of $\partial_{r}I$ except the order of the pole at $r=r_{h}$ do not change
after the MDR was introduced. The corrections to the Hawking temperature were
calculated for massive and charged particles to $\mathcal{O}\left(  m_{p}%
^{-2}\right)  $ and neutral and massless particles to all orders,
respectively. It was found that corrections were suppressed by $m_{p}$.

\item In section \ref{Sec:TOR}, the average number and entropy for a mode were
calculated for bosons and fermions. They could be obtained from those in the
usual case by replacing the standard Hawking temperature with the modified one.

\item In section \ref{Sec:EBW}, we used the brick wall model to compute the
thermal entropy of a massless scalar field near the horizon in UV finite and
perturbative cases. In the UV finite case, the entropy was always finite as
one approached the horizon, and hence the wall near the horizon was not
needed. In the perturbative case, a wall was put at $r=r_{h}+r_{\varepsilon}$
to regulate the UV divergence. We assumed the proper distance between the
horizon and the wall was order of $m_{p}$. Thus, the entropies near the
horizon in both cases were given in eqn. $\left(  \ref{eq:Entropy-MDR}\right)
$. We found that the subleading logarithmic term of the entropy was
independent of the MDR.

\item In section \ref{Sec:BHE}, we calculated luminosities of a 4D spherically
symmetric black hole with the mass $M\gg m_{p}$ and a 2D one. We used the
geometric optics approximation to estimate the effects of scattering off the background.
\end{itemize}

Finally, we briefly discuss the results in this paper and \cite{IN-Wang}. In
this paper and \cite{IN-Wang}, we have calculated the divergent part of the
near horizon atmosphere entropy of a massless scalar field for a 4D
spherically symmetric black hole in the static and free-fall scenarios,
respectively. It appeared that the divergent part in both scenarios could be
presented in the form of a Laurent series with respect to $r_{\varepsilon}$:%
\begin{equation}
S\sim\frac{s_{1}^{0}}{\kappa r_{\varepsilon}}+s_{0}^{0}\ln\kappa
r_{\varepsilon}+%
{\displaystyle\sum\limits_{i=1}^{\infty}}
\frac{T_{0}^{2i}}{m_{p}^{2i}}\left(
{\displaystyle\sum\limits_{j=1}^{\delta_{i}}}
s_{j}^{i}\left(  \kappa r_{\varepsilon}\right)  ^{-j}+s_{0}^{i}\ln\kappa
r_{\varepsilon}\right)  , \label{eq:divergent-4D}%
\end{equation}
where $\delta_{i}=2i+1$ in the static scenario while $\delta_{i}=3i+1$ in the
free-fall scenario.

Assuming the MDR for massless particles is%
\begin{equation}
E^{2}=p%
{\displaystyle\sum\limits_{n=0}}
\tilde{C}_{n}\frac{p^{2n}}{m_{p}^{2n}},
\end{equation}
we found for the emission of $n_{s}$ species of massless scalars and $n_{f}$
species of massless spin-$1/2$ fermions that the total luminosity of a 4D
Schwarzschild black hole was
\begin{equation}
L=\frac{9m_{p}^{2}}{40960M^{2}}\left[  \left(  n_{s}+\frac{7}{4}n_{f}\right)
-\tilde{C}_{1}\left(  0.73n_{s}+1.41n_{f}\right)  \frac{m_{p}^{2}}{M^{2}%
}+\mathcal{O}\left(  \frac{m_{p}^{3}}{M^{3}}\right)  \right]  \text{ in
free-fall frame,} \label{eq:L-ff}%
\end{equation}%
\begin{equation}
L=\frac{9m_{p}^{2}}{40960\pi M^{2}}\left[  \left(  n_{s}+\frac{7}{4}%
n_{f}\right)  +\tilde{C}_{1}\left(  0.48n_{s}+0.92n_{f}\right)  \frac
{m_{p}^{2}}{M^{2}}+\mathcal{O}\left(  \frac{m_{p}^{3}}{M^{3}}\right)  \right]
\text{ in static frame.} \label{eq:L-sf}%
\end{equation}
Note that the sign in front of $\tilde{C}_{1}$ in eqn. $\left(  \ref{eq:L-ff}%
\right)  $ is different from that in eqn. $\left(  \ref{eq:L-sf}\right)  .$
For the sub-luminal dispersion relation with $\tilde{C}_{1}<0$, it means that
the total luminosity increases due to the MDR effects in the free-fall
scenario while it decreases in the static scenario.

\begin{acknowledgments}
We are grateful to Houwen Wu and Zheng Sun for useful discussions. This work
is supported in part by NSFC (Grant No. 11005016, 11175039 and 11375121).
\end{acknowledgments}

\appendix

\section{Effective Field Theory and Deformed Hamilton-Jacobi Equation}

As discussed in the introduction, various approaches to the quantum-gravity
problem could lead to the existence of MDRs. To have a MDR, one has to break
or modify the global Lorentz symmetry in the classical limit of the quantum
gravity. There are several possibilities for breaking or modifying the Lorentz
symmetry, one of which is that Lorentz invariance is spontaneously broken by
extra tensor fields taking on vacuum expectation values. The most conservative
approach for a framework in which to describe MDR is the effective field
theory (EFT), where modifications to the dispersion relation can be described
by the higher dimensional operators. Since we are only interested in
modifications to the dispersion relation of the particles, we limit ourselves
to the kinetic terms and neglect self-interacting effective operators when
constructing the effective field theory. We also assume that the effective
theory respects $U\left(  1\right)  $ gauge invariance of the charged black
hole. The EFT framework can easily incorporate MDR via the introduction of
extra tensors. To construct the minimal EFT in curved spacetime, we suppose
that the action of the EFT contains the usual minimal gravitational couplings
and the EFT coefficients are constants in the local
frame\cite{APD-Kostelecky:2003fs}.

\subsection{Scalar Field}

We work with a complex scalar field $\phi$ with the mass $m$ and the charge
$q$. Following guidelines we put forth, we find the effective Lagrangian for
$\phi$ incorporating MDR can be written as%
\begin{align}
\mathcal{L}_{eff}^{s}  &  =-\phi^{+}\left(  D^{\mu}D_{\mu}+\frac{m^{2}}%
{\hbar^{2}}\right)  \phi-\frac{m^{2}B\left(  \frac{m}{m_{p}}\right)  }%
{\hbar^{2}}\phi^{+}\phi-\frac{imC_{\mu}\left(  \frac{m}{m_{p}}\right)  }%
{\hbar}\phi^{+}D^{\mu}\phi\nonumber\\
&  -%
{\displaystyle\sum\limits_{n\geq2,j}}
\left(  \frac{\hbar}{i}\right)  ^{n-2}\frac{C_{\mu_{1}\cdots\mu_{n}}%
^{j}\left(  \frac{m}{m_{p}}\right)  }{m_{p}^{n-2}}\phi^{+}D^{\mu_{1}}\cdots
D^{\mu_{n}}\phi,
\end{align}
where $D_{\mu}=\nabla_{\mu}+\frac{iq}{\hbar}A_{\mu},$ $\nabla_{\mu}$ is the
covariant derivative of the background spacetime, $A_{\mu\text{ }}$is the
electromagnetic potential, $j$ runs over all independent operators of a given
dimension, $B$ is a dimensionless function of $\frac{m}{m_{p}}$ with $B\left(
0\right)  =0$, and, $C_{\mu}$ and $C_{\mu_{1}\cdots\mu_{n}}^{j}$ are
dimensionless extra tensors depending on $\frac{m}{m_{p}}$ with $C_{\mu
}\left(  0\right)  =C_{\mu\nu}^{j}\left(  0\right)  =0.$ The deformed
Klein-Gordon equation is%
\begin{gather}
-\left(  D^{\mu}D_{\mu}+\frac{m^{2}}{\hbar^{2}}\right)  \phi-\frac
{m^{2}B\left(  \frac{m}{m_{p}}\right)  }{\hbar^{2}}\phi-\frac{imC_{\mu}\left(
\frac{m}{m_{p}}\right)  }{\hbar}D^{\mu}\phi\nonumber\\
-%
{\displaystyle\sum\limits_{n\geq2,j}}
\left(  \frac{\hbar}{i}\right)  ^{n-2}\frac{C_{\mu_{1}\cdots\mu_{n}}%
^{j}\left(  \frac{m}{m_{p}}\right)  }{m_{p}^{n-2}}D^{\mu_{1}}\cdots D^{\mu
_{n}}\phi=0. \label{eq:DeformedKG}%
\end{gather}
With rotational symmetry, all extra tensors become reducible to products of a
vector field $u^{\mu}$, which describes the preferred frame and $u^{\mu}%
u_{\mu}=1$. Thus, the extra tensors become%
\begin{align}
C_{\mu}  &  =C\left(  \frac{m}{m_{p}}\right)  u^{\mu},\nonumber\\
C_{\mu_{1}\cdots\mu_{n}}^{j}  &  =C_{n}^{j}\left(  \frac{m}{m_{p}}\right)
g_{\mu_{i_{1}}\mu_{i_{2}}}\cdots g_{\mu_{i_{2k-1}}\mu_{i_{2k}}}u_{\mu
_{i_{2k+1}}}\cdots u_{\mu_{i_{n}}}, \label{eq:cnj-coef}%
\end{align}
where $g_{\mu\nu}$ is the metric of the background spacetime, $C$ and
$C_{n}^{j}$ are dimensionless functions of $\frac{m}{m_{p}},$ $j=\left(
k,\mathcal{C}\right)  ,$ $2k\leq n,$ and $\mathcal{C}$ denotes any possible
permutations of $\left(  1,\cdots,n\right)  $, namely $\left(  i_{1}%
,\cdots,i_{n}\right)  $. To obtain the Hamilton-Jacobi equation, we make the
WKB ansatz for $\phi$
\begin{equation}
\phi=\exp\left(  \frac{iI}{\hbar}\right)  . \label{eq:ansatzS}%
\end{equation}
Defining
\begin{equation}
\tilde{T}=-u^{\mu}\left(  \partial_{\mu}I+qA_{\mu}\right)  ,\text{ }\tilde
{X}^{2}=\tilde{T}^{2}-\left(  \partial_{\mu}I+qA_{\mu}\right)  ^{2},
\end{equation}
and plugging eqns. $\left(  \ref{eq:ansatzS}\right)  $ and $\left(
\ref{eq:cnj-coef}\right)  $ into eqn. $\left(  \ref{eq:DeformedKG}\right)  $,
one expands eqn. $\left(  \ref{eq:DeformedKG}\right)  $ in powers of $\hbar$
and finds to the lowest order
\begin{equation}
\left(  \tilde{T}^{2}-\tilde{X}^{2}-m^{2}\right)  -m^{2}B\left(  \frac
{m}{m_{p}}\right)  -mC\left(  \frac{m}{m_{p}}\right)  \tilde{T}+%
{\displaystyle\sum\limits_{n\geq2,k\leq\frac{n}{2},\mathcal{C}}}
\frac{\left(  -1\right)  ^{n}C_{n}^{j}\left(  \frac{m}{m_{p}}\right)  \left(
\tilde{T}^{2}-\tilde{X}^{2}\right)  ^{k}\tilde{T}^{n-2k}}{m_{p}^{n-2}}=0.
\label{eq:deformedHJS}%
\end{equation}
Solving eqn. $\left(  \ref{eq:deformedHJS}\right)  $ for $\tilde{X}^{2}$ with
respect to $\tilde{T}$ gives the deformed Hamilton-Jacobi equation for $I$
\begin{equation}
\tilde{X}^{2}=\alpha\left(  \frac{m}{m_{p}}\right)  \tilde{T}^{2}-\beta\left(
\frac{m}{m_{p}}\right)  m^{2}+\gamma\left(  \frac{m}{m_{p}}\right)  m\tilde
{T}+\sum_{n\geq3}\frac{C_{n}\left(  \frac{m}{m_{p}}\right)  \tilde{T}^{n}%
}{m_{p}^{n-2}}, \label{eq:deformedHJSS}%
\end{equation}
where $\alpha,\beta,\gamma$ are dimensionless functions of $\frac{m}{m_{p}}$
with $\alpha\left(  0\right)  =\beta\left(  0\right)  =1$ and $\gamma\left(
0\right)  =0$ which can be determined by the coefficients $B$, $C$ and
$C_{n}^{j}$ in eqn. $\left(  \ref{eq:deformedHJS}\right)  $. In flat spacetime
with $A_{\mu}=0$, the dispersion relation for the scalar field can be found by
inserting the positive energy ansatz $\phi=\exp\left(  -\frac{ip_{\mu}x^{\mu}%
}{\hbar}\right)  $ into eqn. $\left(  \ref{eq:DeformedKG}\right)  $. The
resulting equation for $p_{\mu}$ is actually eqn. $\left(
\ref{eq:deformedHJSS}\right)  $ with $\tilde{T}=u^{\mu}p_{\mu}$ and $\tilde
{X}^{2}=-p_{\mu}p^{\mu}+\tilde{T}^{2}$, which is exact for flat spacetime with
$A_{\mu}=0$. Identifying $E=p_{\mu}u^{\mu}=\tilde{T}$ and $p^{2}=-p_{\mu
}^{\left(  3\right)  }p^{\left(  3\right)  ,\mu}=-p_{\mu}p^{\mu}+\tilde{T}%
^{2}=\tilde{X}^{2}$, we can produce the MDR for the scalar, eqn. $\left(
\ref{eq:MDRExpansion}\right)  ,$ in flat spacetime. On the other hand, the
vector field $u^{\mu}$ is chosen to be $\left(  \frac{1}{\sqrt{f\left(
r\right)  }},\vec{0}\right)  $ in curved spacetime with the metric $\left(
\ref{eq:BHmetric}\right)  $\ and the electromagnetic potential $A_{\mu}$. In
this case, $\tilde{T}$ and $\tilde{X}^{2}$ become $T$ and $X^{2}$ in eqn.
$\left(  \ref{eq:TandX}\right)  $. Thus, in the black hole background
spacetime, the corresponding deformed Hamilton-Jacobi equation for the scalar
field incorporating the MDR, eqn. $\left(  \ref{eq:MDRExpansion}\right)  ,$ is
given by eqn. $\left(  \ref{eq:deformedHJeqn}\right)  .$

\subsection{Fermionic Field}

In the background spacetime with the metric $g_{\mu\nu}$ and the
electromagnetic potential $A_{\mu}$, the effective Lagrangian for a spin-$1/2$
fermion $\psi$ with the mass $m$ and the charge $q$ incorporating the MDR can
be written as%
\begin{gather}
\mathcal{L}_{eff}^{f}=\bar{\psi}\left(  iD_{\mu}^{f}\gamma^{\mu}-\frac
{m}{\hbar}\right)  \psi-\frac{m}{\hbar}%
{\displaystyle\sum\limits_{k\geq0,j}}
B_{\mu_{1}\cdots\mu_{k}}^{j}\left(  \frac{m}{m_{p}}\right)  \bar{\psi}%
\gamma^{\mu_{1}}\cdots\gamma^{\mu_{k}}\psi\nonumber\\
+i%
{\displaystyle\sum\limits_{n\geq k\geq1,j}}
\left(  \frac{\hbar}{i}\right)  ^{k-1}\frac{C_{\mu_{1}\cdots\mu_{n}}%
^{j}\left(  \frac{m}{m_{p}}\right)  }{m_{p}^{k-1}}\bar{\psi}D^{f,\mu_{1}%
}\cdots D^{f,\mu_{k}}\gamma^{\mu_{k+1}}\cdots\gamma^{\mu_{n}}\psi,
\end{gather}
where extra tensors $B_{\mu_{1}\cdots\mu_{k}}^{j}$ and $C_{\mu_{1}\cdots
\mu_{n}}^{j}$ are dimensionless functions of $\frac{m}{m_{p}}$ with
$B_{\mu_{1}\cdots\mu_{k}}^{j}\left(  0\right)  =C_{\mu}^{j}\left(  0\right)
=0$, $j$ runs over all independent operators of a given dimension, $D_{\mu
}^{f}=\partial_{\mu}+\Omega_{\mu}+\frac{iq}{\hbar}A_{\mu}$, $\Omega_{\mu
}\equiv\frac{i}{2}\omega_{\mu}^{\text{ }ab}\Sigma_{ab}$, $\Sigma_{ab}$ is the
Lorentz spinor generator, $\omega_{\mu}^{\text{ }ab}$ is the spin connection
and $\left\{  \gamma_{\mu},\gamma_{\nu}\right\}  =2g_{\mu\nu}$. The Greek
indices are raised and lowered by the curved metric $g_{\mu\nu}$, while the
Latin indices are governed by the flat metric $\eta_{ab}$. The deformed Dirac
equation is%
\begin{gather}
\left(  iD_{\mu}^{f}\gamma^{\mu}-\frac{m}{\hbar}\right)  \psi-\frac{m}{\hbar}%
{\displaystyle\sum\limits_{n\geq0,j}}
B_{\mu_{1}\cdots\mu_{n}}^{j}\left(  \frac{m}{m_{p}}\right)  \gamma^{\mu_{1}%
}\cdots\gamma^{\mu_{n}}\psi\nonumber\\
+i%
{\displaystyle\sum\limits_{n\geq k\geq1,j}}
\left(  \frac{\hbar}{i}\right)  ^{m-1}\frac{C_{\mu_{1}\cdots\mu_{n}}%
^{j}\left(  \frac{m}{m_{p}}\right)  }{m_{p}^{k-1}}D^{f,\mu_{1}}\cdots
D^{f,\mu_{k}}\gamma^{\mu_{k+1}}\cdots\gamma^{\mu_{n}}\psi=0.
\label{eq:deformedDirac}%
\end{gather}
With rotational symmetry, the extra tensors become%
\begin{align}
B_{\mu_{1}\cdots\mu_{n}}^{j}  &  =B_{n}^{j}g_{\mu_{i_{1}}\mu_{i_{2}}}\cdots
g_{\mu_{i_{2k-1}}\mu_{i_{2k}}}u_{\mu_{i_{2k+1}}}\cdots u_{\mu_{i_{n}}%
},\nonumber\\
C_{\mu_{1}\cdots\mu_{n}}^{j}  &  =C_{n}^{j}g_{\mu_{i_{1}}\mu_{i_{2}}}\cdots
g_{\mu_{i_{2k-1}}\mu_{i_{2k}}}u_{\mu_{i_{2k+1}}}\cdots u_{\mu_{i_{n}}},
\end{align}
where $B_{n}^{j}$ and $C_{n}^{j}$ are dimensionless functions of $\frac
{m}{m_{p}},$ $j=\left(  k,\mathcal{C}\right)  ,$ $2k\leq n,$ and $\mathcal{C}$
denotes any possible permutations of $\left(  1,\cdots,n\right)  $, namely
$\left(  i_{1},\cdots,i_{n}\right)  $. To obtain the deformed Hamilton-Jacobi
equation, the ansatz for $\psi$ is assumed as
\begin{equation}
\psi=\exp\left(  \frac{iI}{\hbar}\right)  \mathbf{v}, \label{eq:fermionansatz}%
\end{equation}
where $\mathbf{v}$ is a slowly varying spinor amplitude. Substituting eqn.
$\left(  \ref{eq:fermionansatz}\right)  $ into eqn. $\left(
\ref{eq:deformedDirac}\right)  $, we find to the lowest order of $\hbar$%
\begin{gather}
X_{\mu}\gamma^{\mu}\mathbf{v}+m\mathbf{v}+m%
{\displaystyle\sum\limits_{n\geq0,k\leq\frac{n}{2},\mathcal{C}}}
B_{n}^{j}g_{\mu_{i_{1}}\mu_{i_{2}}}\cdots g_{\mu_{i_{2k-1}}\mu_{i_{2k}}}%
u_{\mu_{i_{2k+1}}}\cdots u_{\mu_{i_{n}}}\gamma^{\mu_{1}}\cdots\gamma^{\mu_{n}%
}\mathbf{v}\nonumber\\
+%
{\displaystyle\sum\limits_{n\geq k\geq1,l\leq\frac{n}{2},\mathcal{C}}}
\frac{C_{n}^{j}}{m_{p}^{k-1}}g_{\mu_{i_{1}}\mu_{i_{2}}}\cdots g_{\mu
_{i_{2l-1}}\mu_{i_{2l}}}u_{\mu_{i_{2l+1}}}\cdots u_{\mu_{i_{n}}}X^{\mu_{1}%
}\cdots X^{\mu_{k}}\gamma^{\mu_{k+1}}\cdots\gamma^{\mu_{n}}\mathbf{v}=0,
\label{eq:deformedHJF}%
\end{gather}
where $X_{\mu}=\partial_{\mu}I+qA_{\mu}$. Using $\left(  X_{\mu}\gamma^{\mu
}\right)  ^{2}=\left(  X_{\mu}\right)  ^{2}=\tilde{T}^{2}-\tilde{X}^{2}$,
$\left(  u_{\mu}\gamma^{\mu}\right)  \left(  X_{\mu}\gamma^{\mu}\right)
=u_{\mu}X^{\mu}=-\tilde{T}$, and $\left(  u_{\mu}\gamma^{\mu}\right)  ^{2}=1,$
one could rewrite eqn. $\left(  \ref{eq:deformedHJF}\right)  $ as%
\begin{equation}
\left[  X_{\mu}+mC_{1}\left(  \frac{m}{m_{p}}\right)  u_{\mu}+h_{2}\left(
\tilde{T},\tilde{X}^{2}\right)  u_{\mu}+g\left(  \tilde{T},\tilde{X}%
^{2}\right)  X_{\mu}\right]  \gamma^{\mu}\mathbf{v}=-\left[  m+mC_{2}\left(
\frac{m}{m_{p}}\right)  +h_{1}\left(  \tilde{T},\tilde{X}^{2}\right)  \right]
\mathbf{v,} \label{eq:deformedHJFF}%
\end{equation}
where $C_{i}$ are dimensionless functions of $\frac{m}{m_{p}}$ with
$C_{i}\left(  0\right)  =0$, $h_{i}\left(  \tilde{T},\tilde{X}^{2}\right)  =%
{\displaystyle\sum\limits_{2p+q\geq1}}
\frac{h_{i}^{p,q}\left(  \frac{m}{m_{p}}\right)  \tilde{T}^{q}\left(
\tilde{T}^{2}-\tilde{X}^{2}\right)  ^{p}}{m_{p}^{2p+q-1}}$ and $g_{i}\left(
\tilde{T},\tilde{X}^{2}\right)  =%
{\displaystyle\sum\limits_{2p+q\geq0}}
\frac{g^{p,q}\left(  \frac{m}{m_{p}}\right)  \tilde{T}^{q}\left(  \tilde
{T}^{2}-\tilde{X}^{2}\right)  ^{p}}{m_{p}^{2p+q}}$. The coefficients $C_{i}$,
$h_{i}^{p,q}$ and $g^{p,q}$ are determined by $B_{n}^{j}$ and $C_{n}^{j}$ from
eqn. $\left(  \ref{eq:deformedHJF}\right)  $. However, the detailed relations
between them are irrelevant here. Multiplying both sides of eqn. $\left(
\ref{eq:deformedHJFF}\right)  $ from the left by $\left(  X_{\mu}+mC_{1}%
u_{\mu}+h_{2}u_{\mu}+gX_{\mu}\right)  \gamma^{\mu}\,$and then using eqn.
$\left(  \ref{eq:deformedHJFF}\right)  $ and $\left\{  \gamma_{\mu}%
,\gamma_{\nu}\right\}  =2g_{\mu\nu}$ to simplify the RHS, one gets%
\begin{equation}
\left(  X^{\mu}+mC_{1}u^{\mu}+h_{2}u^{\mu}+gX^{\mu}\right)  \left(  X_{\mu
}+mC_{1}u_{\mu}+h_{2}u_{\mu}+gX_{\mu}\right)  \mathbf{v}=\left(
m+mC_{2}+h_{1}\right)  ^{2}\mathbf{v}. \label{eq:deformedHJFFF}%
\end{equation}
Since $\mathbf{v}$ is nonzero, eqn. $\left(  \ref{eq:deformedHJFFF}\right)  $
gives
\begin{equation}
\left(  \tilde{T}^{2}-\tilde{X}^{2}-m^{2}\right)  -m^{2}B\left(  \frac
{m}{m_{p}}\right)  -mC\left(  \frac{m}{m_{p}}\right)  \tilde{T}+%
{\displaystyle\sum\limits_{n\geq2,k\leq\frac{n}{2},\mathcal{C}}}
\frac{C_{n}^{j}\left(  \frac{m}{m_{p}}\right)  \left(  \tilde{T}^{2}-\tilde
{X}^{2}\right)  ^{k}\tilde{T}^{n-2k}}{m_{p}^{n-2}}=0,
\label{eq:deformedHJFFFF}%
\end{equation}
where
\begin{gather*}
B=-C_{1}^{2}+C_{2}^{2}+2C_{2},\\
C=2C_{1}+2g^{0,0}C_{1}+2C_{1}h_{2}^{0,1}-2\left(  1+C_{2}\right)  h_{1}%
^{0,1},\\%
{\displaystyle\sum}
\frac{C_{n}^{j}\left(  \tilde{T}^{2}-\tilde{X}^{2}\right)  ^{k}\tilde
{T}^{n-2k}}{m_{p}^{n-2}}=g\left(  \tilde{T}^{2}-\tilde{X}^{2}\right)
+2\left(  h_{2}+gh_{2}\right)  \tilde{T}-h_{1}^{2}+h_{2}^{2}\\
+2m\left(  g-g^{0,0}\right)  C_{1}\tilde{T}+2mC_{1}\left(  h_{2}-h_{2}%
^{0,1}\tilde{T}\right)  -2m\left(  1+C_{2}\right)  \left(  h_{1}-h_{1}%
^{0,1}\tilde{T}\right)  .
\end{gather*}
It is noted that the form of eqn. $\left(  \ref{eq:deformedHJFFF}\right)  $ is
the same as that of eqn. $\left(  \ref{eq:deformedHJS}\right)  $. Thus, the
argument and result below eqn. $\left(  \ref{eq:deformedHJSS}\right)  $ can
also apply to a spin-$1/2$ fermion field.

\end{document}